\documentclass[
reprint,
amsmath,
amssymb,
aps,
]{revtex4-1}

\usepackage[colorlinks, breaklinks=true, linkcolor=blue, citecolor=blue, linktocpage=true]{hyperref}
\usepackage{color}
\usepackage{graphicx}
\usepackage{dcolumn}
\usepackage{bm}

\usepackage{txfonts}

\begin{document}

\title{Valley-Dependent Magnetoresistance in Two-Dimensional Semiconductors}

\author{Akihiko Sekine}
\email[Present address: Center for Emergent Matter Science, RIKEN, Wako, Saitama 351-0198, Japan. ]{akihiko.sekine@riken.jp}
\affiliation{Department of Physics, The University of Texas at Austin, Austin, Texas 78712, USA}
\author{Allan H. MacDonald}
\affiliation{Department of Physics, The University of Texas at Austin, Austin, Texas 78712, USA}

\date{\today}

\begin{abstract}
We show theoretically that two-dimensional direct-gap semiconductors 
with a valley degree of freedom, including 
monolayer transition-metal dichalcogenides and gapped bilayer graphene,
have a longitudinal magnetoconductivity contribution that is
odd in valley and odd in the magnetic field applied perpendicular to the 
system.  Using a quantum kinetic theory we show how this valley-dependent 
magnetoconductivity arises from the interplay between the momentum-space Berry curvature of Bloch 
electrons, the presence of a magnetic field, and disorder scattering.
We discuss how the effect can be measured experimentally
and used as a detector of valley polarization.
\\
\end{abstract}

\maketitle

{\it Introduction.---}
Studies of magnetotransport in metals have a long standing in condensed matter physics.
From the viewpoint of technology the discoveries of giant magnetoresistance \cite{Baibich1988,Binasch1989} 
and tunnel magnetoresistance \cite{Julliere1975,Miyazaki1995,Moodera1995} have led to drastic improvements 
in the performance of magnetic information storage devices.  More generally magnetoresistance
studies can play an important role in characterizing the electronic structure of solids.
For example, Shubnikov--de Haas resistance oscillations are routinely used to 
measure Fermi surfaces.  More recently the existence of three-dimensional (3D) Dirac and Weyl semimetals, which have topologically nontrivial band structures, has been confirmed experimentally \cite{Xiong2015,Li2015,Huang2015,Li2016,Li2016a,Arnold2016} by 
measuring a remarkable and characteristic negative 
longitudinal magnetoresistance property associated with the chiral anomaly \cite{Son2013,Burkov2014,Spivak2016,Sekine2017}.
 
\begin{figure}[!b]
\centering
\includegraphics[width=\columnwidth]{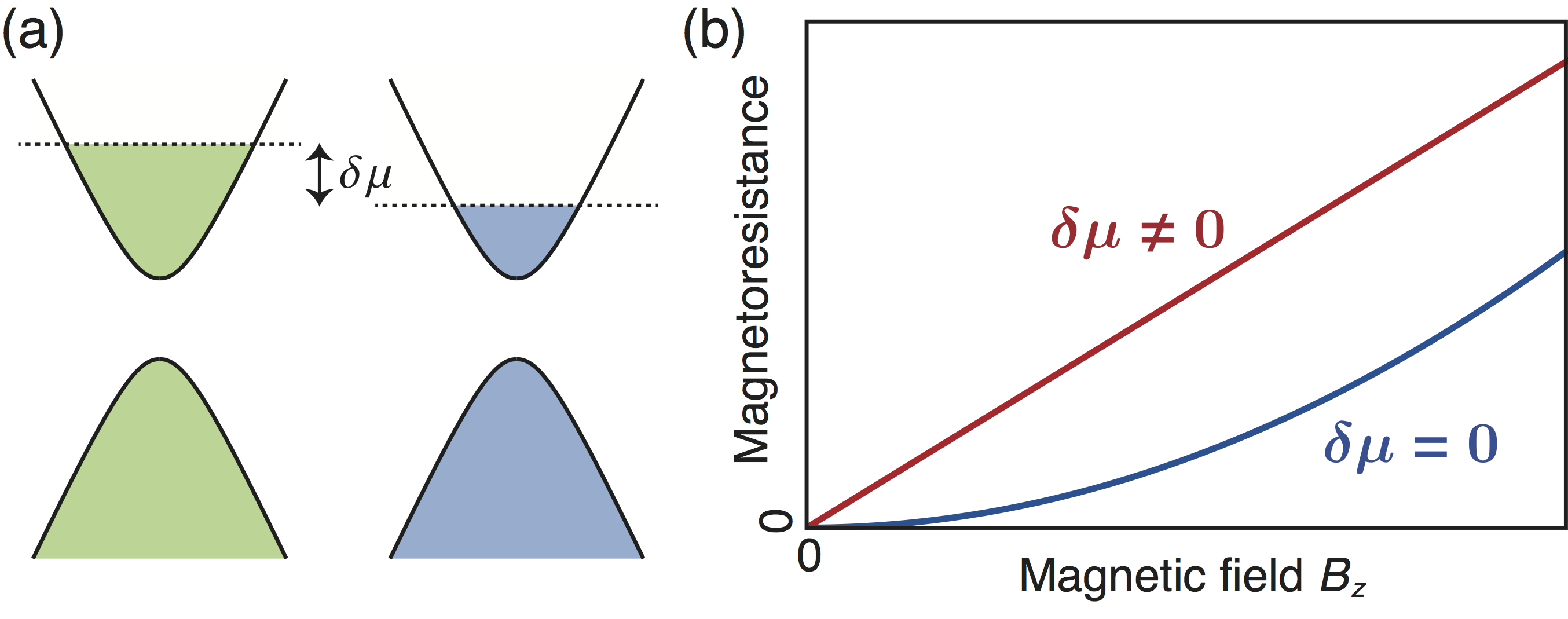}
\caption{(a) Schematic illustration of valley polarization due to  
a chemical potential difference $\delta\mu $ between two valleys
in systems with weak intervalley scattering.
(b) Schematic of the magnetic-field dependence of the
low-magnetic-field magnetoresistance.
The magnetoresistance is predicted to have a linear dependence on $B_z$ when 
$\delta\mu\neq 0$, and a quadratic dependence on $B_z$ when $\delta\mu=0$.
}\label{Fig1}
\end{figure}

This Rapid Communication addresses magnetotransport in 2D semiconductors with more than one 
valley.  Valley has recently attracted greater attention as an 
observable degree of freedom of electrons in solids \cite{Xiao2012,Xu2014,Mak2016},
in part because of the emergence of monolayer transition-metal dichalcogenides (TMDs) and 
gapped bilayer graphene, both 2D semiconductors in which
valence and conduction band extrema occur 
at the $K$ and $K'$ time-reversal partner Brillouin-zone corner points.  
When intervalley scattering by disorder or phonons is weak, valley remains an approximate 
quantum number even beyond the Bloch band approximation.  Weak valley relaxation combined 
with valley-dependent contributions to the conductivity tensor can lead to observable effects 
analogous to those produced by spin accumulations in conductors with weak 
spin-orbit scattering.  To date, attention has 
focused mainly on the valley-dependent anomalous Hall effect \cite{Xiao2007,Mak2014},
which occurs in the absence of a magnetic field 
and is related to the broken time-reversal symmetry of the Hamiltonian's
projection to a single valley, and to momentum-space Berry phase effects.
Given the negative magnetoresistance in 3D Dirac and Weyl semimetals,
which also involves valleys related by time-reversal, 
longitudinal magnetotransport effects should be expected in 2D multi-valley 
systems.  We approach this issue theoretically 
using a massive Dirac model for 2D multi-valley semiconductors 
and a recently developed quantum kinetic theory \cite{Culcer2017,Sekine2017}.
We find that the longitudinal magnetoconductivity has a contribution that is odd in valley and odd in perpendicular 
magnetic field.  Our theoretical predictions can be tested by observing 
a change from quadratic to linear magnetoresistance in 
systems in which a finite valley polarization is induced by optical pumping or 
valley injection, as schematically illustrated in Fig.~\ref{Fig1}.
%

{\it Magnetotransport theory.---}
The transport theory 
we employ is valid in the low magnetic field regime where Landau quantization 
can be neglected and enables us to systematically compute the conductivity tensor in the presence of 
disorder in arbitrary spatial dimensions.  It is based on a quantum kinetic equation that accounts for 
disorder, and for electric $\bm{E}$ and magnetic $\bm{B}$ fields \cite{Sekine2017,Comment6}:
\begin{align}
\frac{\partial \langle \rho\rangle}{\partial t}+\frac{i}{\hbar}[\mathcal{H}_0,\langle \rho\rangle]+K(\langle \rho\rangle)=D_E(\langle \rho\rangle)+D_B(\langle \rho\rangle),
\label{full-kinetic-equation}
\end{align}
where $\langle\rho\rangle$ is the impurity-averaged Bloch-electron density matrix, $\mathcal{H}_0$ is the unperturbed Bloch Hamiltonian, $K(\langle \rho\rangle)$ is a disorder contribution discussed further below,
and $D_E(\langle \rho\rangle)$ and $D_B(\langle \rho\rangle)$ are the electric and magnetic driving terms:
\begin{align}
D_{E}(\langle \rho\rangle)&=\frac{e\bm{E}}{\hbar}\cdot\frac{D\langle \rho\rangle}{D\bm{k}},\nonumber \\
D_{B}(\langle \rho\rangle)&=\frac{e}{2\hbar^2}\left\{\left(\frac{D \mathcal{H}_0}{D\bm{k}}\times\bm{B}\right)\cdot\frac{D\langle \rho\rangle}{D\bm{k}}\right\}.
\label{driving-terms}
\end{align}
Here, $e>0$, $\bm{k}$ is the  crystal wave-vector,
$\{\bm{a}\cdot\bm{b}\}=\bm{a}\cdot\bm{b}+\bm{b}\cdot\bm{a}$ (with $\bm{a}$ and $\bm{b}$ being vectors) 
is a symmetrized operator product, and we have introduced a covariant derivative 
notation for the wave-vector dependence of the matrices
$X$ ($=\langle\rho\rangle,\mathcal{H}_0$) expressed in an eigenstate representation:
\begin{align}
\frac{DX}{D\bm{k}}=\frac{\partial X}{\partial \bm{k}}-i[\bm{\mathcal{R}}_{\bm{k}}, X],
\label{Covariant-derivative}
\end{align}
where $\bm{\mathcal{R}}_{\bm{k}}=\sum_{\alpha=x,y,z}\mathcal{R}_{\bm{k},\alpha}\bm{e}_\alpha$,
and $[\mathcal{R}_{\bm{k},\alpha}]^{mn}=i\langle u^m_{\bm{k}}|\partial_{k_\alpha}u^{n}_{\bm{k}}\rangle$ 
is a generalized Berry connection of Bloch electrons.

The steady-state linear response of the density matrix to an electric field can be expressed as 
a formal expansion in powers of the magnetic field strength $B$ \cite{Sekine2017,Comment5}: 
$\langle\rho\rangle=(1-\mathcal{L}^{-1}D_B)^{-1}\mathcal{L}^{-1}D_{E}(\langle \rho_{0}\rangle+\langle\Xi_B\rangle)\equiv\langle\rho_E\rangle+\sum_{N\ge 1} \langle\rho_{B,N}\rangle$, where $\langle\rho_E\rangle=\mathcal{L}^{-1}D_{E}(\langle \rho_{0}\rangle)$, $\langle \rho_{B,N}\rangle=(\mathcal{L}^{-1}D_B)^N\mathcal{L}^{-1}D_{E}(\langle \rho_{0}\rangle)+(\mathcal{L}^{-1}D_B)^{N-1}\mathcal{L}^{-1}D_{E}(\langle \Xi_B\rangle)$,
and we have defined the Liouvillian operator $\mathcal{L}\equiv P+K$ with 
$P\langle\rho\rangle \equiv (i/\hbar) \, [\mathcal{H}_0,\langle \rho\rangle]$.
In this expansion $\langle\rho_{0}\rangle$ is the Fermi-Dirac equilibrium density matrix in the absence of 
both fields, and $\langle \Xi_B\rangle$ is the equilibrium density matrix in the absence 
of an electric field.  $\langle \Xi_B\rangle$ accounts for the 
Berry phase correction to the density of states implied by semiclassical wave-packet dynamics \cite{Xiao2005}.

Throughout this Rapid Communication, we work in the eigenstate basis for the various contributions to the 
steady-state density matrix, and decompose
$\langle \rho_{B,N}\rangle$ into its band-diagonal part $\langle n_{B,N}\rangle+\langle \xi_{B,N}\rangle$,
and its band-off-diagonal part $\langle S_{B,N}\rangle$.
We adopt a relaxation time approximation for the disorder scattering that influences
the diagonal part of $\langle \rho_{B,N}\rangle$:
\begin{align}
\langle n_{B,N}\rangle_{\bm{k}}^{mm}&= \tau_m [D_B(\langle\rho_{B,N-1}\rangle)]_{\bm{k}}^{mm}, \nonumber\\
\langle \xi_{B,N}\rangle_{\bm{k}}^{mm}&=\frac{e}{\hbar}\bm{B}\cdot\bm{\Omega}_{\bm{k}}^m\, \langle n_{B,N-1}\rangle_{\bm{k}}^{mm},
\label{n_B-Nth-order}
\end{align}
where $N\ge 1$, $\langle\rho_{B,0}\rangle=\langle\rho_E\rangle$, 
$\langle n_{B,0}\rangle=2\langle n_E\rangle$, and $\tau_m$ and $\bm{\Omega}_{\bm{k}}^m$ are
respectively the scattering time and the Berry curvature vector for band $m$.
We also have $\langle \Xi_B\rangle^{mm}_{\bm{k}}=(e/\hbar)\bm{B}\cdot\bm{\Omega}_{\bm{k}}^m \langle\rho_{0}\rangle^{mm}_{\bm{k}}$.
In Eq.~(\ref{n_B-Nth-order}), $\langle n_{B,N}\rangle$ is the extrinsic (Lorentz force) contribution, 
while $\langle \xi_{B,N}\rangle$ is the intrinsic (Berry phase) contribution.
The band off-diagonal part is given by \cite{Sekine2017}
\begin{align}
\langle S_{B,N}\rangle_{\bm{k}}^{mm'}=\frac{\hbar}{i}\frac{[D_B(\langle\rho_{B,N-1}\rangle)]_{\bm{k}}^{mm'}-[J(\langle n_{B,N}\rangle)]_{\bm{k}}^{mm'}}{\varepsilon_{\bm{k}}^m-\varepsilon_{\bm{k}}^{m'}},
\label{S_B-Nth-order}
\end{align}
where $m\neq m'$ and $\varepsilon_{\bm{k}}^m$ is the energy eigenvalue of band $m$.
In Eq.~(\ref{S_B-Nth-order}) the term proportional to $D_B(\langle\rho_{B,N-1}\rangle)$ is purely
a band-structure property expressed in terms of the Berry connection, whereas the term proportional to  
$J(\langle n_{B,N}\rangle)$ is a disorder-dependent Fermi-surface 
response corresponding to a vertex correction in the ladder-diagram approximation \cite{Culcer2017,Sekine2017}.
The explicit form of $J(\langle n_{B,N}\rangle)$ will be given later.

{\it Massive Dirac model.---}
We consider 2D semiconductors with broken inversion symmetry, like monolayer TMDs, 
that have two low-energy valleys related by time-reversal.
The low-energy effective Hamiltonians in these 
systems normally have the massive Dirac form \cite{Xiao2007,Liu2011,Xiao2012}
\begin{align}
\mathcal{H}_{\tau_z}(\bm{k})=v_F (\tau_z k_x\sigma_x+k_y\sigma_y)+m\sigma_z.
\label{Hamiltonian}
\end{align}
(As we discuss briefly later, gated bilayer graphene is an exception.)
In Eq.~(\ref{Hamiltonian}) $\tau_z=\pm 1$ distinguishes the two valleys,
$v_F$ is the Fermi velocity, $2m$ is the band gap, and 
the Pauli matrices $\sigma_i$ act in the space of the retained conduction and valence
bands.  The eigenvalues of the Hamiltonian~(\ref{Hamiltonian}) are 
$\pm\varepsilon_{\bm{k}}=\pm\sqrt{v_F^2(k_x^2+k_y^2)+m^2}$ with the eigenfunctions $|u_{\bm{k}}^\pm(\tau_z)\rangle$.
From Eq.~(\ref{Covariant-derivative}), we see that 
the wavevector dependence of the eigenfunctions $|u_{\bm{k}}^\pm(\tau_z)\rangle$
plays an important role in our transport theory.  
In the eigenstate basis the Berry connection vector 
$[\mathcal{R}^{\tau_z}_{\bm{k},\alpha}]^{mn}=i\langle u^m_{\bm{k}}(\tau_z)|\partial_{k_\alpha}u^{n}_{\bm{k}}(\tau_z)\rangle$ with $m,n=\pm$
has the explicit form 
\begin{align}
\mathcal{R}^{\tau_z}_{\bm{k},x}=\, &\frac{1}{2 k}\tau_z\sin\theta-\tilde{\sigma}_z\frac{m}{2k\varepsilon_{\bm{k}}}\tau_z\sin\theta-\tilde{\sigma}_y\frac{v_F m}{2\varepsilon_{\bm{k}}^2}\cos\theta\nonumber\\
&-\tilde{\sigma}_x\frac{v_F}{2\varepsilon_{\bm{k}}}\tau_z\sin\theta,\nonumber\\
\mathcal{R}^{\tau_z}_{\bm{k},y}=\, &-\frac{1}{2 k}\tau_z\cos\theta+\tilde{\sigma}_z\frac{m}{2k\varepsilon_{\bm{k}}}\tau_z\cos\theta-\tilde{\sigma}_y\frac{v_F m}{2\varepsilon_{\bm{k}}^2}\sin\theta\nonumber\\
&+\tilde{\sigma}_x\frac{v_F}{2\varepsilon_{\bm{k}}}\tau_z\cos\theta,
\end{align}
where $e^{\pm i\theta}=(k_x\pm ik_y)/k$, $k=\sqrt{k_x^2+k_y^2}$, and $\tilde{\sigma}_\alpha$ is 
a Pauli matrix that acts in the eigenstate basis.
Also, the Berry curvature takes the form $[\Omega^{\tau_z}_{\bm{k},z}]^{\pm}=i\langle \partial_{k_x}u_{\bm{k}}^\pm(\tau_z)|\partial_{k_y}u_{\bm{k}}^\pm(\tau_z)\rangle-i\langle \partial_{k_y}u_{\bm{k}}^\pm(\tau_z)|\partial_{k_x}u_{\bm{k}}^\pm(\tau_z)\rangle=\mp\tau_z v_F^2m/(2\varepsilon_{\bm{k}}^3)$.

{\it Valley-dependent longitudinal magnetoconductivity.---}
We apply our magnetotransport theory to the 2D systems described by Eq.~(\ref{Hamiltonian})
with a static magnetic field $\bm{B}=(0,0,B_z)$ applied perpendicular to the system.
For the moment we neglect the vertex 
correction, i.e., the contribution proportional to $J$ in Eq.~(\ref{S_B-Nth-order}).
Without loss of generality we may consider
an electron-doped case with positive chemical potential $\mu$.
Our goal is to compute the $xx$-component of the magnetoconductivity tensor using 
\begin{align}
\sigma^{(N)}_{\mu\nu}(B_z)=\mathrm{Tr}\left[(-e) v_\mu \langle\rho_{B,N}\rangle\right]/E_\nu.
\label{Sigma-xx-formal}
\end{align}
In the eigenstate basis the velocity operator reads
$v_x=v_F(\tilde{\sigma}_z\frac{v_F k}{\varepsilon_{\bm{k}}}\cos\theta+\tilde{\sigma}_y\tau_z\sin\theta-\tilde{\sigma}_x\frac{m}{\varepsilon_{\bm{k}}}\cos\theta)$.

\begin{figure}[!t]
\centering
\includegraphics[width=\columnwidth]{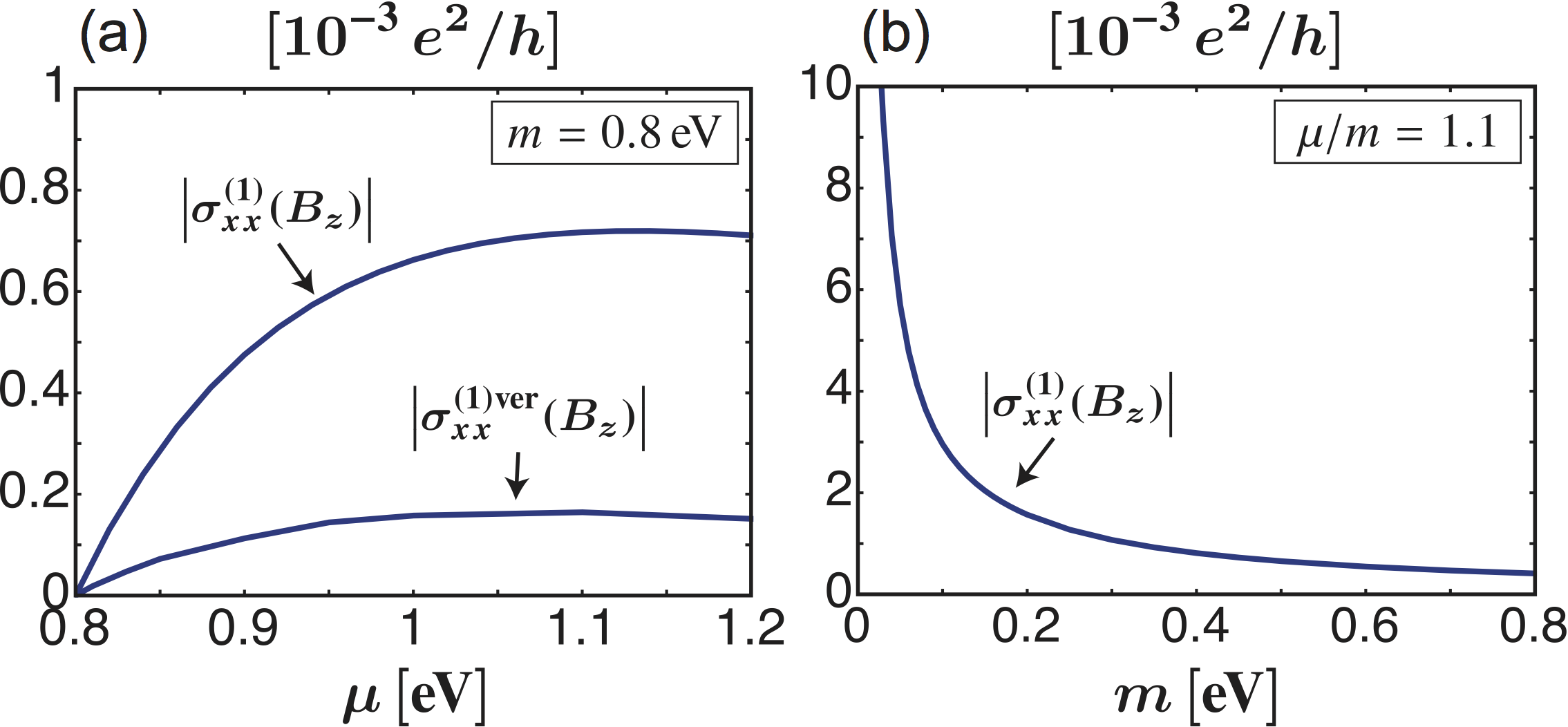}
\caption{
(a) Chemical potential $\mu$ dependence of $\sigma_{xx}^{(1)}(B_z)$ [Eq.~(\ref{MC-Linear})] and $\sigma_{xx}^{(1)\mathrm{ver}}(B_z)$ [Eq.~(\ref{Linear-MC-disorder})] for $m=0.8\, \mathrm{eV}$.
Both $\sigma_{xx}^{(1)}(B_z)$ and $\sigma_{xx}^{(1)\mathrm{ver}}(B_z)$ are proportional 
to $m/\mu^2$ in the case of $\mu \gg m$ and thus approach zero in the limit $\mu \gg m$.
(b) Band gap $m$ dependence of $\sigma_{xx}^{(1)}(B_z)$ for $\mu/m=1.1$.
In both (a) and (b), we set $B_z=0.1\, \mathrm{T}$, $v_F=3\, \mathrm{eV\cdot\AA}$, $\tau_{\mathrm{tr}}=0.1\, \mathrm{ps}$, and $T=5\, \mathrm{meV}$.}
\label{Fig2}
\end{figure}

We first evaluate the magnetoconductivity contributions proportional to odd powers of $B_z$
for $\mu > m$ in the conduction band.
The linear magnetoconductivity $\sigma_{xx}^{(1)}(B_z)$ is determined 
by the density matrix $\langle\rho_{B,1}\rangle=\mathcal{L}^{-1}D_B(\langle \rho_E\rangle)+\mathcal{L}^{-1}D_{E}(\langle \Xi_B\rangle)$.
We find that \cite{Comment1}
\begin{align}
\sigma_{xx}^{(1)}(B_z)&=\tau_z \frac{e^2 B_z}{E_x}\int [d\bm{k}]\left[\frac{v_F^2 m}{2\varepsilon_{\bm{k}}^2}\frac{\partial}{\partial k_x}+\frac{v_F^4 m k_x}{\varepsilon_{\bm{k}}^4}\right]\langle n_E\rangle_{\bm{k}}^{++}\nonumber\\
&\equiv\tau_z \frac{e^3}{\hbar}B_z v_F^2\mathcal{C}_1(\mu,m)\tau_{\mathrm{tr}},
\label{MC-Linear}
\end{align}
where $[d\bm{k}]\equiv\frac{d^2k}{(2\pi)^2}$, $\langle n_E\rangle_{\bm{k}}^{++}=\tau_{\mathrm{tr}}(eE_x/\hbar)\partial f_0(\varepsilon_{\bm{k}})/\partial k_x$, $f_0(\varepsilon_{\bm{k}})$
is the Fermi-Dirac distribution function, $\tau_{\mathrm{tr}}$ is the intravalley
scattering time, and $\mathcal{C}_1(\mu,m)<0$ is evaluated by performing a numerical integration.
The two terms in square brackets of Eq.~(\ref{MC-Linear}) acting on the extrinsic response $\langle n_E\rangle$ arise respectively from the off-diagonal intrinsic contribution $\langle S_{B,1}\rangle$ and the diagonal intrinsic contribution $\langle \xi_{B,1}\rangle$.
In Fig.~\ref{Fig2} we show the $\mu$ and $m$ dependences of $\sigma_{xx}^{(1)}(B_z)$.
It should be noted here that, as seen from Fig.~\ref{Fig2}(b), $\sigma_{xx}^{(1)}(B_z)$ is significantly enhanced in a smaller gap system.
Similarly, we find the cubic magnetoconductivity obtained from $\langle\rho_{B,3}\rangle$ \cite{Comment1}
\begin{align}
\sigma^{(3)}_{xx}(B_z)&=\frac{8}{3}\tau_z \frac{e^2 B_z}{E_x}\int [d\bm{k}]\left[\frac{v_F^2 m}{2\varepsilon_{\bm{k}}^2}\frac{\partial}{\partial k_x}+\frac{v_F^4 m k_x}{2\varepsilon_{\bm{k}}^4}\right]\langle n_{B,2}\rangle_{\bm{k}}^{++}\nonumber\\
&\equiv\tau_z \frac{e^5}{\hbar}B_z^3 v_F^6\mathcal{C}_3(\mu,m)\tau_{\mathrm{tr}}^3,
\label{MC-Cubic}
\end{align}
where $\langle n_{B,2}\rangle_{\bm{k}}^{++}=(eB_z\tau_{\mathrm{tr}})^2(\frac{\partial \varepsilon_{\bm{k}}}{\partial k_y}\frac{\partial}{\partial k_x}-\frac{\partial \varepsilon_{\bm{k}}}{\partial k_x}\frac{\partial}{\partial k_y})^2 \langle n_E\rangle_{\bm{k}}^{++}$, and $\mathcal{C}_3(\mu,m)>0$.
In the case of $\mu\gg m$, we find that $\mathcal{C}_3(\mu,m)\propto m/\mu^4$.
For the material parameters used in Fig.~\ref{Fig2}(a), 
$|\sigma^{(3)}_{xx}(B_z)/\sigma^{(1)}_{xx}(B_z)| \sim 10^{-3}(B_z\, [\rm{T}])^2$.
The two terms in square brackets of Eq.~(\ref{MC-Cubic}) acting on the extrinsic response $\langle n_{B,2}\rangle$ arise respectively from the off-diagonal intrinsic contribution $\langle S_{B,3}\rangle$ and the diagonal intrinsic contribution $\langle \xi_{B,3}\rangle$.
There are no valley-independent contributions to the linear and cubic 
magnetoconductivities, as required by time-reversal symmetry.
Higher-order odd-power terms have the 
general form
\begin{align}
\sigma^{(N)}_{xx}(B_z)=\tau_z \frac{e^{N+2}}{\hbar} B_z^N v_F^{2N}\mathcal{C}_N(\mu,m)\tau_{\mathrm{tr}}^N,
\end{align}
where $N=5,7,9\cdots$ is an odd integer and $\mathcal{C}_N(\mu,m)$ has the dimension of [Energy]$^{-N}$.

Next, we consider the magnetoconductivity contributions proportional to even powers of $B_z$,
which cannot be valley dependent due to time-reversal symmetry \cite{Comment2}.  
We find that the quadratic magnetoconductivity obtained from $\langle\rho_{B,2}\rangle$ 
is dominated by the Lorentz-force contribution
\begin{align}
\sigma^{(2)}_{xx}(B_z)
&\approx -\frac{e^3 B_z^2\tau_{\mathrm{tr}}^2}{E_x}\int [d\bm{k}]\frac{v_F^2 k_x}{\varepsilon_{\bm{k}}}\biggl(\frac{\partial \varepsilon_{\bm{k}}}{\partial k_y}\frac{\partial}{\partial k_x}-\frac{\partial \varepsilon_{\bm{k}}}{\partial k_x}\frac{\partial}{\partial k_y}\biggr)^2 \langle n_E\rangle_{\bm{k}}^{++} \nonumber\\
&= -\sigma_{xx}^{(0)}(\omega_c\tau_{\mathrm{tr}})^2,
\label{quadratic-MC}
\end{align}
where $\sigma_{xx}^{(0)}=(-e/E_x)\int [d\bm{k}](v_F^2 k_x/\varepsilon_{\bm{k}})\langle n_E\rangle_{\bm{k}}^{++}$ is the Drude conductivity and $\omega_c=eB_zv_F^2/\mu$ is the cyclotron frequency.
Intrinsic contributions to the quadratic magnetoconductivity are not zero, but they are
suppressed by $\sim 1/(\mu\tau_{\mathrm{tr}})^2\ll 1$ 
compared to the conventional contribution in Eq.~(\ref{quadratic-MC}) \cite{Comment1}.

{\it Vertex corrections.---}
From Eq.~(\ref{S_B-Nth-order}) the vertex correction contribution to the density matrix linear in $B_z$ is given by $\langle S'_{B,1}\rangle^{mm''}_{\bm{k}}\equiv i\hbar[J(\langle n_{B,1}\rangle)]^{mm''}_{\bm{k}}/(\varepsilon_{\bm{k}}^m-\varepsilon_{\bm{k}}^{m''})$, where \cite{Culcer2017}
\begin{align}
[J(\langle n\rangle)]^{mm''}_{\bm{k}} =&\ \frac{\pi}{\hbar} \sum_{m'\bm{k}'} \langle U^{mm'}_{\bm{k}\bm{k}'}U^{m'm''}_{\bm{k}'\bm{k}}\rangle \left[ (n^{m}_{\bm{k}} - n^{m'}_{\bm{k}'})\delta(\varepsilon^m_{\bm{k}} - \varepsilon^{m'}_{\bm{k}'}) \right. \nonumber\\
&\left. +\, (n^{m''}_{\bm{k}} - n^{m'}_{\bm{k}'})\delta(\varepsilon^{m''}_{\bm{k}} - \varepsilon^{m'}_{\bm{k}'})\right].
\label{Anomalous-driving-term}
\end{align}
Here, $m \ne m''$ and $\langle n\rangle=\mathrm{diag}[n^{m}_{\bm{k}}]$ is an arbitrary band-diagonal density matrix.
We assume short-range disorder of the form $U(\bm{r})=U_0\sum_i\delta(\bm{r}-\bm{r}_i)$
with $\langle U(\bm{r})U(\bm{r}')\rangle=n_{\mathrm{imp}}U_0^2\,\delta(\bm{r}-\bm{r}')$, 
where $n_{\mathrm{imp}}$ is the impurity density.
After a lengthy calculation \cite{Comment1}, we find the vertex correction to the linear magnetoconductivity 
\begin{align}
\sigma_{xx}^{(1)\mathrm{ver}}(B_z)&=\mathrm{Tr}\left[(-e) v_x \langle S'_{B,1}\rangle\right]/E_x \nonumber\\
&\equiv\tau_z \frac{e^3}{\hbar}B_z v_F^2\mathcal{C}_1^{\mathrm{ver}}(\mu,m)\tau_{\mathrm{tr}},
\label{Linear-MC-disorder}
\end{align}
where $\mathcal{C}_1^{\mathrm{ver}}(\mu,m)<0$.
This means that the vertex correction 
enhances the valley-dependent linear magnetoconductivity [see Fig.~\ref{Fig2}(a)].
This contrasts with its well-known influence on the 
the spin Hall \cite{Inoue2004} and anomalous Hall \cite{Inoue2006} conductivities in certain Rashba models, i.e., the suppression of these conductivities by the vertex correction.
Here, we note that usual golden-rule intraband scattering rates proportional to
 $\langle U^{++}_{\bm{k}\bm{k}'}U^{++}_{\bm{k}'\bm{k}}\rangle$ are not valley dependent.
Nonzero contributions to Eq.~(\ref{Linear-MC-disorder}) require interband scattering
matrix elements like $\langle U^{++}_{\bm{k}\bm{k}'}U^{+-}_{\bm{k}'\bm{k}}\rangle$,
which are valley dependent.  The vertex correction is valley dependent because it is due to  
interband coherence induced by the magnetic field.

{\it Discussion.---}
A longitudinal total magnetoconductivity proportional to odd powers of magnetic field can
occur only in systems with broken time-reversal symmetry  \cite{Onsager1931,Chen2015}.
In monolayer TMDs and gated bilayer graphene, spatial inversion symmetry
is broken but time-reversal symmetry is retained.
The valley-dependent magnetoconductivity contributes to transport only 
in the presence of a finite valley polarization, for example one due to a 
chemical potential difference between the two valleys, that explicitly breaks time-reversal symmetry.
Valley polarization in TMDs can be realized by applying circularly polarized 
light \cite{Xiao2012,Cao2012,Zeng2012,Mak2012} to generate an excess population of 
carriers in one valley.  When intravalley scattering is much stronger than intervalley scattering, 
equilibration will occur within valleys to establish valley-dependent chemical potentials.
This approach has been used previously to 
measure the valley Hall effect \cite{Mak2014},
and is the most direct way to measure the valley-dependent magnetoconductivity
derived in this Rapid Communication.
In discussing the results of such a measurement below, 
we assume that the contribution to transport from photo-generated holes is
negligible.

\begin{figure}[!t]
\centering
\includegraphics[width=\columnwidth]{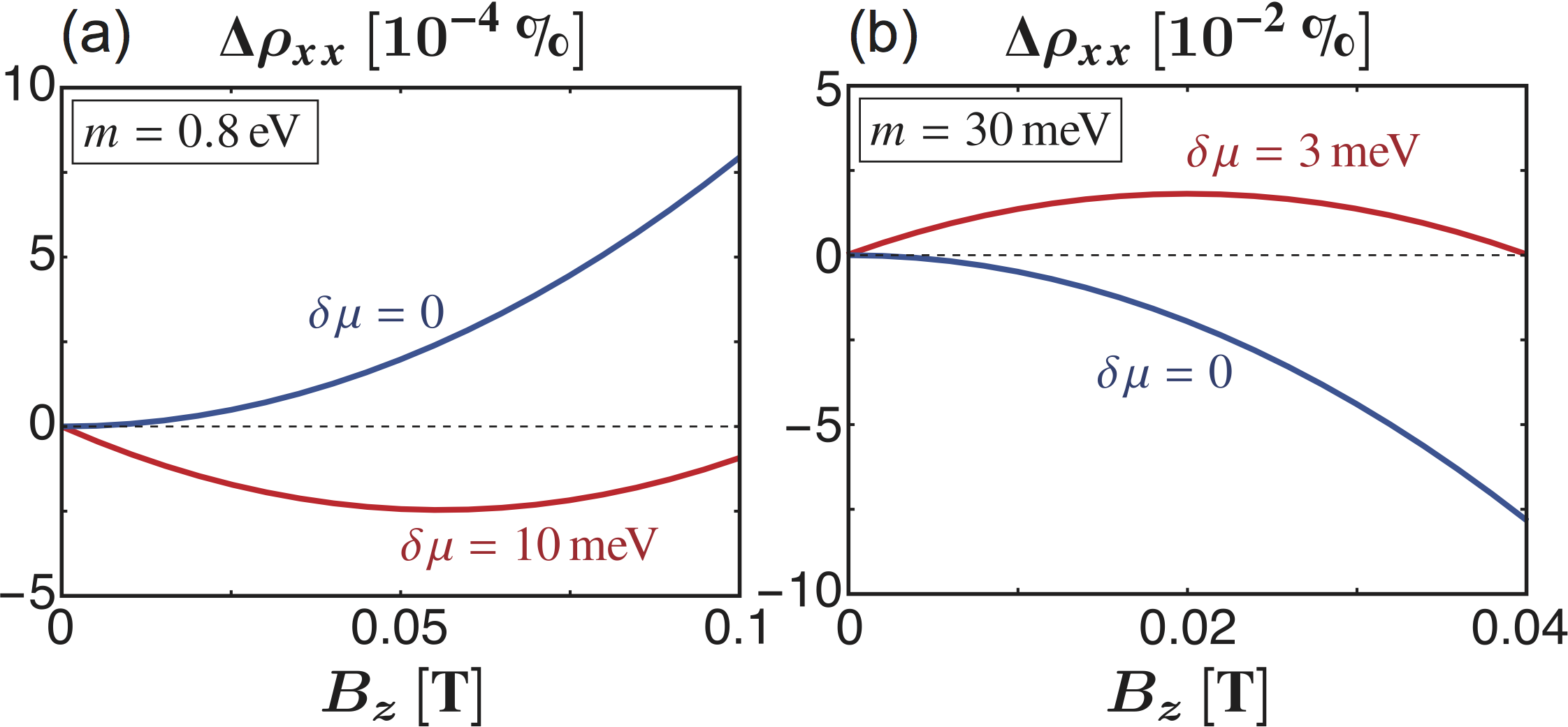}
\caption{
(a) Valley polarization $\delta\mu$ dependence of the magnetoresistance $\Delta\rho_{xx}(B_z,\delta\mu)$  [Eq.~(\ref{Magnetoresistance})] in the case of $|\sigma_{xx}^{(0)}/\sigma_{xy}^{(0)}|\gg 1$ for $m=0.8\, \mathrm{eV}$, corresponding to the typical gap of monolayer TMDs.
We used $\mu_1=\mu_2=0.82\, \mathrm{eV}$ for the $\delta\mu=0$ case, and 
$\mu_1=0.83\, \mathrm{eV}$ and $\mu_2=0.82\, \mathrm{eV}$ for the $\delta\mu\neq0$ case.
(b) $\Delta\rho_{xx}(B_z,\delta\mu)$ in the case of $|\sigma_{xx}^{(0)}/\sigma_{xy}^{(0)}|\ll 1$ for $m= 30\, \mathrm{meV}$, corresponding to low-carrier-density gated bilayer graphene.
We used $\mu_1=\mu_2=33\, \mathrm{meV}$ for the $\delta\mu=0$ case, and $\mu_1=36\, \mathrm{meV}$ and $\mu_2=33\, \mathrm{meV}$ for the $\delta\mu\neq0$ case.
Note that the magnetoresistance effect 
is much stronger in a smaller gap system.
In both (a) and (b), we set $v_F=3\, \mathrm{eV\cdot\AA}$, $\tau_{\mathrm{tr}}=0.1\, \mathrm{ps}$, and $T=5\, \mathrm{meV}$.
}\label{Fig3}
\end{figure}

Including terms up to order of $B_z^2$ and allowing for a chemical potential 
difference between valleys, the total magnetoconductivity of a two-valley system reads
\begin{align}
\sigma_{xx}^{B}(\mu_1,\mu_2)=\, &\frac{e^3}{\hbar}B_z v_F^2[\mathcal{C}^{\mathrm{tot}}_1(\mu_1,m)-\mathcal{C}^{\mathrm{tot}}_1(\mu_2,m)]\tau_{\mathrm{tr}} \nonumber\\
&-\sigma_{xx}^{(0)}(\mu_1)(\omega_{c1}\tau_{\mathrm{tr}})^2 - \sigma_{xx}^{(0)}(\mu_2)(\omega_{c2}\tau_{\mathrm{tr}})^2,
\label{sigma_xx-full}
\end{align}
where $\mu_i$ ($i=1,2$) is the chemical potential of valley 
$i$, $\mathcal{C}^{\mathrm{tot}}_1(\mu_i,m)=\mathcal{C}_1(\mu_i,m)+\mathcal{C}^{\mathrm{ver}}_1(\mu_i,m)$, 
and $\omega_{ci}=eB_zv_F^2/\mu_i$.  In the low-field limit 
$\sigma_{xx}^{B}\propto B_z$ when $\delta\mu=\mu_1-\mu_2\neq 0$, while 
$\sigma_{xx}^{B}\propto B_z^2$ when $\delta\mu=0$.
The resistivity is defined by $\rho_{xx}(B_z)=\sigma_{xx}/(\sigma_{xx}^2+\sigma_{xy}^2)$ 
with $\sigma_{\mu\nu}\equiv\sigma_{\mu\nu}^{(0)}+\sigma_{\mu\nu}^{B}$.
It follows that the low-field magnetoresistance 
\begin{equation}
\Delta\rho_{xx} \equiv  \frac{\rho_{xx}(B_z)-\rho_{xx}(0)}{\rho_{xx}(0)} 
\approx 
\mp\frac{\sigma_{xx}^{B}(\mu_1,\mu_2)}{\sigma_{xx}^{(0)}(\mu_1)+\sigma_{xx}^{(0)}(\mu_2)}.
\label{Magnetoresistance}
\end{equation}
Here, the $-$ ($+$) sign applies in the $|\sigma_{xx}^{(0)}/\sigma_{xy}^{(0)}|\gg 1$ case ($|\sigma_{xx}^{(0)}/\sigma_{xy}^{(0)}|\ll 1$ case) \cite{Comment3}.
Obviously the Drude conductivity $\sigma_{xx}^{(0)}$ is not valley dependent.
Thus the change in magnetic-field dependence from $B_z^2$ to $B_z$ when illuminated by circularly polarized light,
illustrated schematically in Fig.~\ref{Fig1},
should be readily observable \cite{Comment4}.
Interestingly, we can always make the $\delta\mu\neq 0$ magnetoresistance in the low-field limit 
opposite in sign to the $\delta\mu=0$ magnetoresistance by 
changing the sense of circular light polarization, as illustrated in Fig.~\ref{Fig3}.

The magnetoresistance effects discussed in this Rapid Communication are much stronger, for a given Fermi velocity,
in Dirac models with a smaller gap, and we expect them to be much more easily observed 
experimentally in bilayer graphene systems than in monolayer TMDs.
Bilayer graphene is described by a generalized Dirac model with chirality $J=2$ rather than $J=1$, and 
has quadratic 
dispersion in the absence of a gap \cite{Min2008}.  For a given gap the size of the 
magnetoresistance effect in bilayer graphene will exceed the $J=1$ model
values plotted in Fig.~\ref{Fig3}(b). 

A valley-dependent conductivity can lead to valley accumulation in the absence of 
optical valley pumping when an inhomogeneity is present along the current path, for 
example a variation in carrier density induced by external gates.  Because valley dependence is 
largest in a relative sense when the carrier density is small (i.e., when the Fermi energy 
$\mu$ is only slightly larger than the gap $m$), the current partitioning between valleys corresponding 
to equal electro-chemical potential gradients changes across interfaces at which the carrier 
density changes.  If intervalley scattering is weak, valley accumulation will persist within a 
valley relaxation length of any such interface, and should be detectable via Kerr microscopy \cite{Lee2016}.

{\it Summary.---}
To summarize, we have theoretically demonstrated the existence of a
 valley-dependent longitudinal magnetoconductivity in 2D semiconductors with a valley degree of 
 freedom.  The effect arises from the interplay between the momentum-space Berry curvature of Bloch 
electrons, the presence of a magnetic field, and disorder scattering.
Our prediction can be verified by measuring the influence of circularly-polarized light illumination 
on magnetoresistance in the low-field limit.
Because the magnetoresistance is proportional to valley polarization it 
can be used as a valley detector.
We predict that these magnetoresistance effects
will be significantly enhanced in bilayer graphene samples with small gaps and small carrier densities.

\vspace{1.5ex}
We thank H. Chen and D. Culcer for valuable discussions.
This work was supported by the Department of Energy, Office of Basic Energy Sciences under Contract No. DE-FG02-ER45958 and by the Welch foundation under Grant No. TBF1473.
A.S. is supported by the JSPS Overseas Research Fellowship.


\setcounter{figure}{0}
\setcounter{equation}{0}
\setcounter{table}{0}
\renewcommand{\thefigure}{S\arabic{figure}}
\renewcommand{\theequation}{S\arabic{equation}}
\renewcommand{\thetable}{S\Roman{table}}

\begin{widetext}
\vspace{4ex}
\begin{center}
\textbf{{\Large Supplemental Material}}
\end{center}
\section{Steady-state linear response of the density matrix to an electric field}
Let us consider a density matrix in the presence of electric and magnetic fields.
We write the electron density matrix as $\langle\rho\rangle=\langle\rho_0\rangle+\langle\rho\rangle_F$, where $\langle\rho_0\rangle$ is the Fermi-Dirac equilibrium density matrix in the absence of fields, and $\langle\rho\rangle_F$ is the field-induced density matrix.
Then we can rewrite the steady-state uniform system limit of Eq.~(1) in the form
\begin{equation} 
(\mathcal{L} - D_E - D_B)\langle\rho\rangle_F = (D_E+D_B)\langle \rho_0\rangle,
\end{equation}
where we have defined the Liouvillian operator $\mathcal{L}\equiv P + K$ with $P\langle\rho\rangle \equiv (i/\hbar) \, [\mathcal{H}_0,\langle \rho\rangle]$ and have used the fact that $\mathcal{L}\langle \rho_0\rangle=0$.
It follows that
\begin{align}
\langle\rho\rangle_F&=\bigl[1-\mathcal{L}^{-1}(D_E+D_{B})\bigr]^{-1} \mathcal{L}^{-1}(D_E+D_B)\langle \rho_0\rangle \nonumber\\
&=\sum_{N\ge 0} \bigl[\mathcal{L}^{-1}(D_E+D_B)\bigr]^N \mathcal{L}^{-1}(D_E+D_B)\langle \rho_0\rangle.
\label{rho-expansion-general}
\end{align}
Equation~(\ref{rho-expansion-general}) can be used to derive a low-magnetic-field expansion for the linear response 
of the density matrix, and hence of any single-particle observable, to an electric field $\bm{E}$.
From Eq.~(\ref{rho-expansion-general}) we have
\begin{align}
\langle\rho\rangle_F=\sum_{N,N'\ge 0} (\mathcal{L}^{-1}D_B)^N \mathcal{L}^{-1}D_{E} (\mathcal{L}^{-1}D_B)^{N'}\langle \rho_0\rangle.
\end{align}
Here, the $N=N'=0$ term is the density matrix induced solely by the electric field $\bm{E}$ (which is linear in $\bm{E}$): $\langle\rho_E\rangle=\mathcal{L}^{-1}D_{E}(\langle \rho_{0}\rangle)$.
It is convenient to define a density matrix induced solely by the magnetic field $\bm{B}$ as $\langle \rho_B\rangle\equiv\sum_{N\ge 1}(\mathcal{L}^{-1}D_B)^{N}\langle \rho_0\rangle$.
Note that this expression is the generalized solution for $(\mathcal{L} - D_{B}) \langle\rho_B\rangle = D_B(\langle\rho_0\rangle)$.
Since we are assuming that the magnetic field is very weak, we retain only the term linear in $\bm{B}$ and neglect the higher-order terms, i.e., we may set $\langle \rho_B\rangle=\langle\Xi_B\rangle$, where $\langle\Xi_B\rangle$ accounts for the Berry phase correction to the density of states implied by semiclassical wave-packet dynamics.
This means that the correction to the Fermi-Dirac distribution function due to $\bm{B}$ is given by the Berry phase correction $\langle\Xi_B\rangle$.
Finally, we obtain
\begin{align}
\langle\rho\rangle_F=\langle\rho_E\rangle+\sum_{N\ge 1} \langle\rho_{B,N}\rangle,
\end{align}
where $\langle \rho_{B,N}\rangle=(\mathcal{L}^{-1}D_B)^N\mathcal{L}^{-1}D_{E}(\langle \rho_{0}\rangle)+(\mathcal{L}^{-1}D_B)^{N-1}\mathcal{L}^{-1}D_{E}(\langle \Xi_B\rangle)$.

\section{General expression for the magnetic driving term}
Let us consider a generic two-band system with the energy eigenvalues $\varepsilon^\pm_{\bm{k}}=\pm \varepsilon_{\bm{k}}$ and a magnetic field applied along the $z$ direction such that $\bm{B}=(0,0,B_z)$.
In what follows we work in the eigenstate basis where matrices are written in the basis of $\left[\begin{array}{cc} ++ & +- \\ -+ & --\end{array}\right]$.
The magnetic driving term obtained from a diagonal density matrix $\langle n\rangle=\left[\begin{array}{cc} n^+_{\bm{k}} & 0 \\ 0 & n^-_{\bm{k}}\end{array}\right]$ is given by
\begin{align}
D_B(\langle n\rangle)&=\frac{1}{2}eB_z\left[\left\{\frac{D\mathcal{H}_0}{Dk_y},\frac{D\langle n\rangle}{Dk_x}\right\}-\left\{\frac{D\mathcal{H}_0}{Dk_x},\frac{D\langle n\rangle}{Dk_y}\right\}\right] \nonumber\\
&=eB_z
\begin{bmatrix}
\frac{\partial \varepsilon_{\bm{k}}}{\partial k_y}\frac{\partial n^+_{\bm{k}}}{\partial k_x}-\frac{\partial \varepsilon_{\bm{k}}}{\partial k_x}\frac{\partial n^+_{\bm{k}}}{\partial k_y} && -i\varepsilon_{\bm{k}}(\mathcal{R}_{x}^{+-}\frac{\partial}{\partial k_y}-\mathcal{R}_{y}^{+-}\frac{\partial}{\partial k_x})(n^+_{\bm{k}}+n^-_{\bm{k}})\\
i\varepsilon_{\bm{k}}(\mathcal{R}_{x}^{-+}\frac{\partial}{\partial k_y}-\mathcal{R}_{y}^{-+}\frac{\partial}{\partial k_x})(n^+_{\bm{k}}+n^-_{\bm{k}}) && -\left(\frac{\partial \varepsilon_{\bm{k}}}{\partial k_y}\frac{\partial n^-_{\bm{k}}}{\partial k_x}-\frac{\partial \varepsilon_{\bm{k}}}{\partial k_x}\frac{\partial n^-_{\bm{k}}}{\partial k_y}\right)
\end{bmatrix}.
\label{D_B-from-diagonal-density}
\end{align}
On the other hand, the magnetic driving term obtained from an off-diagonal density matrix $\langle S\rangle=\left[\begin{array}{cc} 0 & a_{\bm{k}} \\ b_{\bm{k}} & 0\end{array}\right]$ is given by
\begin{align}
D_B(\langle S\rangle)=\frac{1}{2}eB_z\left[\left\{\frac{D\mathcal{H}_0}{Dk_y},\frac{D\langle S\rangle}{Dk_x}\right\}-\left\{\frac{D\mathcal{H}_0}{Dk_x},\frac{D\langle S\rangle}{Dk_y}\right\}\right]=eB_z
\begin{bmatrix}
\mathcal{A}_{\bm{k}}-\mathcal{B}_{\bm{k}} && 0\\
0 &&\mathcal{A}_{\bm{k}}-\mathcal{B}_{\bm{k}}
\end{bmatrix}
\label{D_B-from-offdiagonal-density}
\end{align}
with
\begin{align}
\left\{
\begin{aligned}
\mathcal{A}_{\bm{k}}&=-i\frac{\partial \varepsilon_{\bm{k}}}{\partial k_y}(\mathcal{R}_x^{+-}b_{\bm{k}}-\mathcal{R}_x^{-+}a_{\bm{k}})+i\mathcal{R}_y^{+-}\varepsilon_{\bm{k}}\left[i(\mathcal{R}_x^{++}-\mathcal{R}_x^{--})b_{\bm{k}}+\frac{\partial b_{\bm{k}}}{\partial k_x}\right] -i\mathcal{R}_y^{-+}\varepsilon_{\bm{k}}\left[-i(\mathcal{R}_x^{++}-\mathcal{R}_x^{--})a_{\bm{k}}+\frac{\partial a_{\bm{k}}}{\partial k_x}\right],\\
\mathcal{B}_{\bm{k}}&=-i\frac{\partial \varepsilon_{\bm{k}}}{\partial k_x}(\mathcal{R}_y^{+-}b_{\bm{k}}-\mathcal{R}_y^{-+}a_{\bm{k}})+i\mathcal{R}_x^{+-}\varepsilon_{\bm{k}}\left[i(\mathcal{R}_y^{++}-\mathcal{R}_y^{--})b_{\bm{k}}+\frac{\partial b_{\bm{k}}}{\partial k_y}\right] -i\mathcal{R}_x^{-+}\varepsilon_{\bm{k}}\left[-i(\mathcal{R}_y^{++}-\mathcal{R}_y^{--})a_{\bm{k}}+\frac{\partial a_{\bm{k}}}{\partial k_y}\right].
\end{aligned}
\right.
\end{align}
Especially in the case of $\langle S\rangle=c_{1\bm{k}}\sigma_y$ (i.e., $a_{\bm{k}}=-ic_{1\bm{k}}$ and $b_{\bm{k}}=ic_{1\bm{k}}$), we have
\begin{align}
\left\{
\begin{aligned}
\mathcal{A}_{\bm{k}}&=(\mathcal{R}_x^{+-}+\mathcal{R}_x^{-+})\frac{\partial \varepsilon_{\bm{k}}}{\partial k_y}c_{1\bm{k}}-i(\mathcal{R}_y^{+-}-\mathcal{R}_y^{-+})(\mathcal{R}_x^{++}-\mathcal{R}_x^{--})\varepsilon_{\bm{k}}c_{1\bm{k}}-(\mathcal{R}_y^{+-}+\mathcal{R}_y^{-+})\varepsilon_{\bm{k}}\frac{\partial c_{1\bm{k}}}{\partial k_x},\\
\mathcal{B}_{\bm{k}}&=(\mathcal{R}_y^{+-}+\mathcal{R}_y^{-+})\frac{\partial \varepsilon_{\bm{k}}}{\partial k_x}c_{1\bm{k}}-i(\mathcal{R}_x^{+-}-\mathcal{R}_x^{-+})(\mathcal{R}_y^{++}-\mathcal{R}_y^{--})\varepsilon_{\bm{k}}c_{1\bm{k}}-(\mathcal{R}_x^{+-}+\mathcal{R}_x^{-+})\varepsilon_{\bm{k}}\frac{\partial c_{1\bm{k}}}{\partial k_y}.
\end{aligned}\label{Expression-of-AB-sigma_y}
\right.
\end{align}
Similarly in the case of $\langle S\rangle=c_{2\bm{k}}\sigma_x$ (i.e., $a_{\bm{k}}=c_{2\bm{k}}$ and $b_{\bm{k}}=c_{2\bm{k}}$), we have
\begin{align}
\left\{
\begin{aligned}
\mathcal{A}_{\bm{k}}&=-i(\mathcal{R}_x^{+-}-\mathcal{R}_x^{-+})\frac{\partial \varepsilon_{\bm{k}}}{\partial k_y}c_{2\bm{k}}-(\mathcal{R}_y^{+-}+\mathcal{R}_y^{-+})(\mathcal{R}_x^{++}-\mathcal{R}_x^{--})\varepsilon_{\bm{k}}c_{2\bm{k}}+i(\mathcal{R}_y^{+-}-\mathcal{R}_y^{-+})\varepsilon_{\bm{k}}\frac{\partial c_{2\bm{k}}}{\partial k_x},\\
\mathcal{B}_{\bm{k}}&=-i(\mathcal{R}_y^{+-}-\mathcal{R}_y^{-+})\frac{\partial \varepsilon_{\bm{k}}}{\partial k_x}c_{2\bm{k}}-(\mathcal{R}_x^{+-}+\mathcal{R}_x^{-+})(\mathcal{R}_y^{++}-\mathcal{R}_y^{--})\varepsilon_{\bm{k}}c_{2\bm{k}}+i(\mathcal{R}_x^{+-}-\mathcal{R}_x^{-+})\varepsilon_{\bm{k}}\frac{\partial c_{2\bm{k}}}{\partial k_y}.
\end{aligned}\label{Expression-of-AB-sigma_x}
\right.
\end{align}

\section{Theoretical Model}
Let us consider the two-component massive Dirac fermion model whose Hamiltonian is given by
\begin{align}
\mathcal{H}_{\tau_z}(\bm{k})=v_F(\tau_z k_x\sigma_x+k_y\sigma_y)+m\sigma_z
\label{MassiveDirac}
\end{align}
with $\tau_z=\pm 1$ being the valley index.
The eigenvectors are given by
\begin{align}
|u_{\bm{k}}^\pm(\tau_z)\rangle&=\frac{1}{\sqrt{2}}
\begin{bmatrix}
\sqrt{1\pm\frac{m}{\varepsilon_{\bm{k}}}}\\
\pm \tau_z e^{i\tau_z\theta}\sqrt{1\mp\frac{m}{\varepsilon_{\bm{k}}}}
\end{bmatrix},
\end{align}
where $\varepsilon^\pm_{\bm{k}}=\pm\varepsilon_{\bm{k}}=\pm\sqrt{v_F^2(k_x^2+k_y^2)+m^2}$ are the eigenvalues and $e^{\pm i\theta}=(k_x\pm ik_y)/k$ with $k=\sqrt{k_x^2+k_y^2}$.
The Berry connection $[\mathcal{R}^{\tau_z}_{\bm{k},\alpha}]^{mn}=i\langle u_{\bm{k}}^m(\tau_z)|\frac{\partial}{\partial k_\alpha}u_{\bm{k}}^n(\tau_z)\rangle$ with $m,n=\pm$
has the explicit form
\begin{align}
\mathcal{R}_{\bm{k},x}^{\tau_z}&=\frac{1}{2 k}\tau_z\sin\theta-\tilde{\sigma}_z\frac{m}{2k\varepsilon_{\bm{k}}}\tau_z\sin\theta-\tilde{\sigma}_y\frac{v_F m}{2\varepsilon_{\bm{k}}^2}\cos\theta-\tilde{\sigma}_x\frac{v_F}{2\varepsilon_{\bm{k}}}\tau_z\sin\theta,\nonumber\\
\mathcal{R}_{\bm{k},y}^{\tau_z}&=-\frac{1}{2 k}\tau_z\cos\theta+\tilde{\sigma}_z\frac{m}{2k\varepsilon_{\bm{k}}}\tau_z\cos\theta-\tilde{\sigma}_y\frac{v_F m}{2\varepsilon_{\bm{k}}^2}\sin\theta+\tilde{\sigma}_x\frac{v_F}{2\varepsilon_{\bm{k}}}\tau_z\cos\theta.
\end{align}
Also, the Berry curvature takes the form $[\Omega^{\tau_z}_{\bm{k},z}]^{\pm}=i\langle \partial_{k_x}u_{\bm{k}}^\pm(\tau_z)|\partial_{k_y}u_{\bm{k}}^\pm(\tau_z)\rangle-i\langle \partial_{k_y}u_{\bm{k}}^\pm(\tau_z)|\partial_{k_x}u_{\bm{k}}^\pm(\tau_z)\rangle=\mp\tau_z v_F^2m/(2\varepsilon_{\bm{k}}^3)$.
The velocity operator in the eigenstate representation is obtained as
\begin{align}
v_x(\tau_z)=\langle u_{\bm{k}}^m(\tau_z)|\frac{\partial \mathcal{H}_{\tau_z}}{\partial k_x}|u_{\bm{k}}^n(\tau_z)\rangle
=v_F\left(\tilde{\sigma}_z\frac{v_F k}{\varepsilon_{\bm{k}}}\cos\theta+\tilde{\sigma}_y\tau_z\sin\theta-\tilde{\sigma}_x\frac{m}{\varepsilon_{\bm{k}}}\cos\theta\right).
\label{v_x}
\end{align}

\section{Linear magnetoconductivity}
Let us consider a case where an electric field is applied along the $x$ direction and a magnetic field is applied along the $z$ direction such that $\bm{E}=(E_x,0,0)$ and $\bm{B}=(0,0,B_z)$.
The longitudinal linear magnetoconductivity is given by
\begin{align}
\sigma^{(1)}_{xx}(B_z)=\mathrm{Tr}\left[(-e) v_x \left\{\mathcal{L}^{-1}D_B (\langle\rho_E\rangle)+\mathcal{L}^{-1}D_E(\langle\Xi_B\rangle)\right]\right\}/E_x,
\end{align}
where $\langle\rho_E\rangle=\mathcal{L}^{-1}D_E(\langle\rho_0\rangle)$ is the density matrix linear in the electric field, and $\langle\Xi_B\rangle^{mm}_{\bm{k}}=(e/\hbar)f_0(\varepsilon_{\bm{k}}^m)\bm{B}\cdot\bm{\Omega}^m_{\bm{k}}$ is the density matrix linear in the magnetic field, which corresponds to the Berry phase correction to the density of states in semiclassical wave-packet dynamics.
Here, $\langle\rho_0\rangle=\mathrm{diag}[f_0(\varepsilon_{\bm{k}}^+),f_0(\varepsilon_{\bm{k}}^-)]$ with $\mu>0$ the chemical potential is the Fermi-Dirac distribution function.

We start from the diagonal part of $\langle\rho_E\rangle$, which is given by
\begin{align}
\langle n_E\rangle=\tau_{\mathrm{tr}}
\begin{bmatrix}
eE_x\frac{\partial f_0(\varepsilon_{\bm{k}}^+)}{\partial k_x} &&0\\
0 && 0
\end{bmatrix},
\end{align}
where $\tau_{\mathrm{tr}}$ is the intravalley scattering time.
Next we compute the magnetic driving term obtained from $\langle n_E\rangle$.
From Eq.~(\ref{D_B-from-diagonal-density}) we have
\begin{align}
D_B(\langle n_E\rangle)=eB_z
\begin{bmatrix}
\frac{\partial \varepsilon_{\bm{k}}}{\partial k_y}\frac{\partial n^+_{E\bm{k}}}{\partial k_x}-\frac{\partial \varepsilon_{\bm{k}}}{\partial k_x}\frac{\partial n^+_{E\bm{k}}}{\partial k_y} && -i\varepsilon_{\bm{k}}\left(\mathcal{R}_{x}^{+-}\frac{\partial n^+_{E\bm{k}}}{\partial k_y}-\mathcal{R}_{y}^{+-}\frac{\partial n^+_{E\bm{k}}}{\partial k_x}\right)\\
i\varepsilon_{\bm{k}}\left(\mathcal{R}_{x}^{-+}\frac{\partial n^+_{E\bm{k}}}{\partial k_y}-\mathcal{R}_{y}^{-+}\frac{\partial n^+_{E\bm{k}}}{\partial k_x}\right) && 0
\end{bmatrix}.
\end{align}
Then the corresponding diagonal and off-diagonal parts of the density matrix $\langle \rho_{B}\rangle(=\langle S_{B}\rangle+\langle n_{B}\rangle+\langle\xi_B\rangle)$ are obtained as
\begin{align}
\langle S_{B}\rangle&=\mathcal{L}^{-1}[D_B(\langle n_E\rangle)]=-i\hbar\frac{[D_B(\langle n_E\rangle)]_{\bm{k}}^{mm'}}{\varepsilon_{\bm{k}}^m-\varepsilon_{\bm{k}}^{m'}}
=\frac{eB_z}{2}\left\{
\frac{\partial n^+_{E\bm{k}}}{\partial k_x}
\begin{bmatrix}
0 && \mathcal{R}^{+-}_{\bm{k},y}\\
\mathcal{R}^{-+}_{\bm{k},y}\ && 0
\end{bmatrix}
-\frac{\partial n^+_{E\bm{k}}}{\partial k_y}
\begin{bmatrix}
0 && \mathcal{R}^{+-}_{\bm{k},x}\\
\mathcal{R}^{-+}_{\bm{k},x} && 0
\end{bmatrix}
\right\}\nonumber\\
&=\frac{eB_z}{2}\left\{
\tilde{\sigma}_y\left(-\frac{\partial n^+_{E\bm{k}}}{\partial k_x}\frac{v_F m}{2\varepsilon_{\bm{k}}^2}\sin\theta+\frac{\partial n^+_{E\bm{k}}}{\partial k_y}\frac{v_F m}{2\varepsilon_{\bm{k}}^2}\cos\theta\right)
+\tilde{\sigma}_x\left(\frac{\partial n^+_{E\bm{k}}}{\partial k_x}\frac{\tau_z v_F}{2\varepsilon_{\bm{k}}}\cos\theta+\frac{\partial n^+_{E\bm{k}}}{\partial k_y}\frac{\tau_z v_F}{2\varepsilon_{\bm{k}}}\sin\theta\right)
\right\},
\label{S_EB}
\end{align}
and
\begin{align}
\langle n_{B}\rangle=\mathcal{L}^{-1}[D_B(\langle n_E\rangle)]=\tau_{\mathrm{tr}}[D_B(\langle n_E\rangle)]_{\bm{k}}^{mm}=\tau_{\mathrm{tr}}eB_z
\begin{bmatrix}
\frac{\partial \varepsilon_{\bm{k}}}{\partial k_y}\frac{\partial n^+_{E\bm{k}}}{\partial k_x}-\frac{\partial \varepsilon_{\bm{k}}}{\partial k_x}\frac{\partial n^+_{E\bm{k}}}{\partial k_y} &&0\\
0 && 0
\end{bmatrix}.
\label{n_EB}
\end{align}
Note that $\langle S_{B}\rangle$ is accompanied by the Berry curvature contribution term that is purely diagonal:
\begin{align}
\langle\xi_B\rangle^{++}_{\bm{k}}=eB_z\Omega^+_{\bm{k},z} n_{E\bm{k}}^{+}=- \tau_z eB_z\frac{v_F^2m}{2\varepsilon_{\bm{k}}^3}n^+_{E\bm{k}},\ \ \ \ \ \ \ \ \ \ \langle\xi_B\rangle^{--}_{\bm{k}}=0.
\end{align}
The evaluation of $\mathcal{L}^{-1}D_E(\langle\Xi_B\rangle)$ is quite similar to that of $\langle n_E\rangle$, which gives $\mathcal{L}^{-1}D_E(\langle\Xi_B\rangle)=\langle\xi_B\rangle$.
Note that the off-diagonal part of $\mathcal{L}^{-1}D_E(\langle\Xi_B\rangle)$ does not contribute to the linear magnetoconductivity.

Finally, the total longitudinal linear magnetoconductivity is calculated to be
\begin{align}
\sigma_{xx}^{(1)}(B_z)&=\mathrm{Tr}\left[(-e) v_x (\langle S_{B}\rangle+2\langle\xi_B\rangle)\right]/E_x
=\tau_z \frac{e^2 B_z}{E_x}\int\frac{d^2k}{(2\pi)^2}\left[\frac{v_F^2 m}{2\varepsilon_{\bm{k}}^2}\frac{\partial}{\partial k_x}+\frac{v_F^4 m k_x}{\varepsilon_{\bm{k}}^4}\right] n_{E\bm{k}}^+\nonumber\\
&\equiv\tau_z \frac{e^3}{\hbar}B_z v_F^2\mathcal{C}_1(\mu,m)\tau_{\mathrm{tr}},
\label{MC-Linear-appendix}
\end{align}
where $\mathcal{C}_1(\mu,m)<0$.
Note that the contribution from $\langle n_{B}\rangle$ vanishes because it is an odd function of $k_y$, and that the contribution from $\langle n'_{B}\rangle$ in Fig.~\ref{FigS1} also vanishes because $\langle n'_{B}\rangle$ is proportional to the identity matrix.
In the case of $\mu\gg m$, we find that $\mathcal{C}_1(\mu,m)\propto m/\mu^2$.
Using this relation, we can check that the dimension of $\sigma_{xx}^{(1)}(B_z)$ is indeed the dimension of two-dimensional electrical conductivity:
\begin{align}
\tau_z \frac{e^3}{\hbar}B_z v_F^2\frac{m}{\mu^2}\tau_{\mathrm{tr}}=\mathrm{(A\cdot s)^3\ \frac{1}{J\cdot s}\ \frac{V\cdot s}{m^2}\ \frac{m^2}{s^2}\ \frac{1}{J}\ s}=\mathrm{\frac{A^3s^2V}{J^2}}=\mathrm{\frac{A}{V}},
\end{align}
where we have used $\mathrm{J=V\cdot A\cdot s}$ and $\hbar=\mathrm{J\cdot s}$.

\section{Quadratic magnetoconductivity}
\begin{figure}[!t]
\centering
\includegraphics[width=0.78\columnwidth]{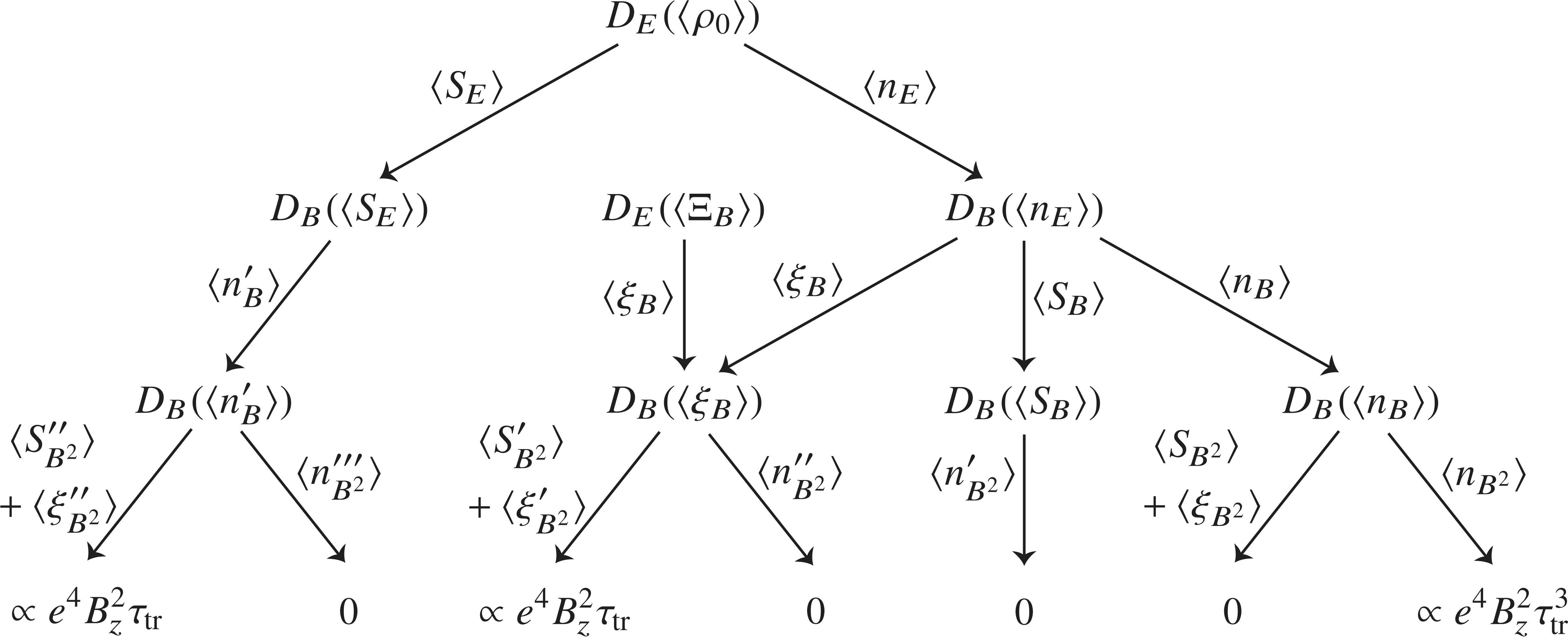}
\caption{Schematic illustration of longitudinal quadratic magnetoconductivity contributions.
$\langle n\rangle$ and $\langle \xi\rangle$ indicates band-diagonal density matrix components, and $\langle S\rangle$ indicates band-off-diagonal density matrix components (see the main text).
Note that, for a generic two-band model, the magnetic driving terms acting on purely off-diagonal density matrices are purely diagonal and proportional to the identity matrix [see Eq.~(\ref{D_B-from-offdiagonal-density})].
}\label{FigS1}
\end{figure}
Let us calculate the quadratic magnetoconductivity, which is given by
\begin{align}
\sigma^{(2)}_{xx}(B_z)=\mathrm{Tr}\left\{(-e) v_x \left[(\mathcal{L}^{-1}D_B)^2\mathcal{L}^{-1}D_E(\langle\rho_0\rangle)+\mathcal{L}^{-1}D_B\mathcal{L}^{-1}D_E(\langle\Xi_B\rangle)\right]\right\}/E_x.
\end{align}
First we consider the contribution from the Lorentz force.
This contribution is purely extrinsic, i.e., comes from the diagonal part of $D_B$.
From Eq.~(\ref{D_B-from-diagonal-density}) the diagonal part of the density matrix obtained from $\langle n_{B}\rangle$ reads
\begin{align}
\langle n_{B^2}\rangle=\mathcal{L}^{-1}[D_B(\langle n_{B}\rangle)]=\tau_{\mathrm{tr}}eB_z
\begin{bmatrix}
\frac{\partial \varepsilon_{\bm{k}}}{\partial k_y}\frac{\partial n^+_{B\bm{k}}}{\partial k_x}-\frac{\partial \varepsilon_{\bm{k}}}{\partial k_x}\frac{\partial n^+_{B\bm{k}}}{\partial k_y} &&0\\
0 && 0
\end{bmatrix}.
\label{n-B^2}
\end{align}
Then we immediately get
\begin{align}
\sigma^{(2)\mathrm{LF}}_{xx}(B_z)&=
-\frac{e^3 B_z^2\tau_{\mathrm{tr}}^2}{E_x}\int \frac{d^2k}{(2\pi)^2}\frac{v_F^2 k_x}{\varepsilon_{\bm{k}}}\biggl(\frac{\partial \varepsilon_{\bm{k}}}{\partial k_y}\frac{\partial}{\partial k_x}-\frac{\partial \varepsilon_{\bm{k}}}{\partial k_x}\frac{\partial}{\partial k_y}\biggr)^2 n_{E\bm{k}}^{+}\nonumber\\
&\equiv -\sigma_{xx}^{(0)}(\omega_c\tau_{\mathrm{tr}})^2,
\label{quadratic-MC}
\end{align}
where $\sigma_{xx}^{(0)}=-e/E_x\int \frac{d^2k}{(2\pi)^2}(v_F^2 k_x/\varepsilon_{\bm{k}})n_{E\bm{k}}^{+}$ is the Drude conductivity and $\omega_c=eB_zv_F^2/\mu$ is the cyclotron frequency.

There also exist intrinsic contributions to the quadratic magnetoconductivity (see Fig.~\ref{FigS1}).
After a calculation we find that the intrinsic contributions take the form
\begin{align}
\sigma^{(2)\mathrm{in}}_{xx}(B_z)= e^4B_z^2v_F^4\mathcal{C}_2(\mu,m)\tau_{\mathrm{tr}},
\end{align}
which is not dependent on $\tau_z$.
We numerically find that $\mathcal{C}_2(\mu,m)\propto m/\mu^4$.
Then we see that $\sigma^{(2)\mathrm{in}}_{xx}(B_z)$ is quite small compared to $\sigma^{(2)\mathrm{LF}}_{xx}(B_z)$:
\begin{align}
\frac{\sigma^{(2)\mathrm{in}}_{xx}(B_z)}{\sigma^{(2)\mathrm{LF}}_{xx}(B_z)}\sim \frac{e^4B_z^2v_F^4(m/\mu^4)\tau_{\mathrm{tr}}}{e^4B_z^2v_F^4(1/\mu)\tau_{\mathrm{tr}}^3}
\sim \frac{1}{(\mu\tau_{\mathrm{tr}})^2}\frac{m}{\mu}\ll 1,
\end{align}
where we have used the fact that the condition $(\mu\tau_{\mathrm{tr}})^2\gg 1$ is usually satisfied in semiconductors.
Therefore the total longitudinal quadratic magnetoconductivity is given by
\begin{align}
\sigma^{(2)}_{xx}(B_z)\approx \sigma^{(2)\mathrm{LF}}_{xx}(B_z)=-\sigma_{xx}^{(0)}(\omega_c\tau_{\mathrm{tr}})^2.
\end{align}
Note that there are no $\tau_z$-dependent contributions to $\sigma^{(2)}_{xx}(B_z)$, as required by time-reversal symmetry.

\section{Cubic magnetoconductivity}
Let us calculate the cubic magnetoconductivity, which is given by
\begin{align}
\sigma^{(3)}_{xx}(B_z)=\mathrm{Tr}\left\{(-e) v_x \left[(\mathcal{L}^{-1}D_B)^3\mathcal{L}^{-1}D_E(\langle\rho_0\rangle)+(\mathcal{L}^{-1}D_B)^2\mathcal{L}^{-1}D_E(\langle\Xi_B\rangle)\right]\right\}/E_x.
\end{align}
The contributions to $\sigma^{(3)}_{xx}(B_z)$ are obtained from the magnetic driving terms acting on the density matrices of the order of $B_z^2$ that are shown in Fig.~\ref{FigS1}.

The off-diagonal density matrix $\langle S_{B^3}\rangle$ obtained from the off-diagonal part of $D_B(\langle n_{B^2}\rangle)$ is given by [see Eq.~(\ref{S_EB}) for a similar calculation]
\begin{align}
\langle S_{B^3}\rangle&=\mathcal{L}^{-1}D_B(\langle n_{B^2}\rangle)=\frac{eB_z}{2}\left\{
\tilde{\sigma}_y\left(-\frac{\partial n^+_{B^2\bm{k}}}{\partial k_x}\frac{v_F m}{2\varepsilon_{\bm{k}}^2}\sin\theta+\frac{\partial n^+_{B^2\bm{k}}}{\partial k_y}\frac{v_F m}{2\varepsilon_{\bm{k}}^2}\cos\theta\right)
+\tilde{\sigma}_x\left(\frac{\partial n^+_{B^2\bm{k}}}{\partial k_x}\frac{\tau_z v_F}{2\varepsilon_{\bm{k}}}\cos\theta+\frac{\partial n^+_{B^2\bm{k}}}{\partial k_y}\frac{\tau_z v_F}{2\varepsilon_{\bm{k}}}\sin\theta\right)
\right\},
\end{align}
which is accompanied by the Berry curvature contribution term
\begin{align}
\langle\xi_{B^3}\rangle^{++}_{\bm{k}}=eB_z\Omega^+_z\langle n_{B^2}\rangle^{++}_{\bm{k}}=-\tau_z eB_z\frac{v_F^2m}{2\varepsilon_{\bm{k}}^3}n^+_{B^2\bm{k}},\ \ \ \ \ \ \ \ \ \ \langle\xi_{B^3}\rangle^{--}_{\bm{k}}=0,
\end{align}
where $n^+_{B^2\bm{k}}$ is given by Eq.~(\ref{n-B^2}).
Then one of the contributions to the cubic magnetoconductivity is calculated to be
\begin{align}
\sigma^{(3)}_{xx}(B_z)&=\mathrm{Tr}\left[(-e) v_x (\langle S_{B^3}\rangle+\langle\xi_{B^3}\rangle)\right]/E_x=
\tau_z eB_z\int_{-\infty}^\infty\frac{d^2k}{(2\pi)^2}\left[\frac{m v_F^2}{2\varepsilon_{\bm{k}}^2}\frac{\partial }{\partial k_x}+\frac{v_F^4m k_x}{2\varepsilon_{\bm{k}}^4}\right]n^+_{B^2\bm{k}}\nonumber\\
&\equiv\tau_z \frac{e^5}{\hbar}B_z^3 v_F^6\mathcal{D}(\mu,m)\tau_{\mathrm{tr}}^3,
\label{cubic-MC2}
\end{align}
where $\mathcal{D}(\mu,m)>0$.
In the case of $\mu\gg m$, we numerically find that $\mathcal{D}(\mu,m)\propto m/\mu^4$.

There is another contribution to the cubic magnetoconductivity.
The diagonal density matrix $\langle n'_{B^2}\rangle$ obtained from $\langle S_{B}\rangle=\mathcal{L}^{-1}D_B(\langle n_E\rangle)=c_1\tilde{\sigma}_y+c_2\tilde{\sigma}_x$ [Eq.~(\ref{S_EB})] is calculated from Eq.~(\ref{D_B-from-offdiagonal-density}) to be
\begin{align}
\langle n'_{B^2}\rangle=\mathcal{L}^{-1}[D_B(\langle S_{B}\rangle)]=\tau_{\mathrm{tr}}eB_z\mathcal{G}_{\bm{k}}\bm{1},
\end{align}
where
\begin{align}
\mathcal{G}_{\bm{k}}=-\tau_z\frac{v_F^3 k}{\varepsilon_{\bm{k}}^2}c_{1\bm{k}}-\tau_z\frac{v_F m^2}{\varepsilon_{\bm{k}}^2 k}c_{1\bm{k}}-\tau_z v_F\cos\theta\frac{\partial c_{1\bm{k}}}{\partial k_x}-\tau_z v_F\sin\theta\frac{\partial c_{1\bm{k}}}{\partial k_y}-\frac{v_F m}{\varepsilon_{\bm{k}}}\sin\theta\frac{\partial c_{2\bm{k}}}{\partial k_x}+\frac{v_F m}{\varepsilon_{\bm{k}}}\cos\theta\frac{\partial c_{2\bm{k}}}{\partial k_y},
\end{align}
with
\begin{align}
c_{1\bm{k}}&=\frac{eB_z}{2}\left(-\frac{\partial n^+_{E\bm{k}}}{\partial k_x}\frac{v_F m}{2\varepsilon_{\bm{k}}^2}\sin\theta+\frac{\partial n^+_{E\bm{k}}}{\partial k_y}\frac{v_F m}{2\varepsilon_{\bm{k}}^2}\cos\theta\right),\\
c_{2\bm{k}}&=\frac{eB_z}{2}\left(\frac{\partial n^+_{E\bm{k}}}{\partial k_x}\frac{\tau_z v_F}{2\varepsilon_{\bm{k}}}\cos\theta+\frac{\partial n^+_{E\bm{k}}}{\partial k_y}\frac{\tau_z v_F}{2\varepsilon_{\bm{k}}}\sin\theta\right).
\end{align}
Here, note that all the terms in $\mathcal{G}_{\bm{k}}$ are dependent on $\tau_z$.
Then, the diagonal density matrix $\langle n'_{B^3}\rangle$ obtained from the diagonal part of $D_B(\langle n'_{B^2}\rangle)$ is written as
\begin{align}
\langle n'_{B^3}\rangle=\mathcal{L}^{-1}[D_B(\langle n'_{B^2}\rangle)]=(eB_z)^2\tau_{\mathrm{tr}}\left(\frac{\partial \varepsilon_{\bm{k}}}{\partial k_y}\frac{\partial \mathcal{G}_{\bm{k}}}{\partial k_x}-\frac{\partial \varepsilon_{\bm{k}}}{\partial k_x}\frac{\partial \mathcal{G}_{\bm{k}}}{\partial k_y}\right)\tilde{\sigma}_z,
\end{align}
from which the contribution to the cubic magnetoconductivity is calculated to be
\begin{align}
\sigma^{(3)}_{xx}(B_z)&=\mathrm{Tr}\left[(-e) v_x \langle n'_{B^3}\rangle\right]/E_x=
-2e^3B_z^2\tau_{\mathrm{tr}}\int_{-\infty}^\infty\frac{d^2k}{(2\pi)^2}\frac{v_F^2 k_x}{\varepsilon_{\bm{k}}}\left(\frac{\partial \varepsilon_{\bm{k}}}{\partial k_y}\frac{\partial \mathcal{G}_{\bm{k}}}{\partial k_x}-\frac{\partial \varepsilon_{\bm{k}}}{\partial k_x}\frac{\partial \mathcal{G}_{\bm{k}}}{\partial k_y}\right)\nonumber\\
&=\frac{2}{3}\tau_z \frac{e^5}{\hbar}B_z^3v_F^6\mathcal{D}(\mu,m)\tau_{\mathrm{tr}}^3.
\label{cubic-MC1}
\end{align}

The other two contributions come from the diagonal parts of the density matrices at each power of $B_z$ acting on $\langle n_E\rangle$: one of the three actions of $D_B$ is the Berry curvature contribution $eB_z\Omega^+_z$, and the other two actions of $D_B$ are the Lorentz force contribution $\tau_{\mathrm{tr}}eB_z[\frac{\partial \varepsilon_{\bm{k}}}{\partial k_y}\frac{\partial }{\partial k_x}-\frac{\partial \varepsilon_{\bm{k}}}{\partial k_x}\frac{\partial }{\partial k_y}]$.
After a calculation we find that the other contribution to the cubic magnetoconductivity is
\begin{align}
\sigma^{(3)}_{xx}(B_z)=\tau_z \frac{e^5}{\hbar}B_z^3 v_F^6\mathcal{D}(\mu,m)\tau_{\mathrm{tr}}^3.
\label{cubic-MC3}
\end{align}

Finally, from Eqs.~(\ref{cubic-MC2}), (\ref{cubic-MC1}), and (\ref{cubic-MC3}) we obtain the total longitudinal cubic magnetoconductivity
\begin{align}
\sigma^{(3)}_{xx}(B_z)=\tau_z \frac{e^5}{\hbar}B_z^3 v_F^6\mathcal{C}_3(\mu,m)\tau_{\mathrm{tr}}^3,
\label{cubic-MC-total}
\end{align}
where $\mathcal{C}_3(\mu,m)=(8/3)\mathcal{D}(\mu,m)<0$.
Note that there are no $\tau_z$-independent contributions to $\sigma^{(3)}_{xx}(B_z)$, as required by time-reversal symmetry.

\section{Disorder Effect in the Born Approximation}
Let us consider a short-range (on-site) disorder of the form $U(\bm{r})=U_0\sum_i\delta(\bm{r}-\bm{r}_i)$, and assume that the correlation function satisfies $\langle U(\bm{r})U(\bm{r}')\rangle=n_{\mathrm{imp}}U_0^2\,\delta(\bm{r}-\bm{r}')$ with $n_{\mathrm{imp}}$ the impurity density.
The eigenvectors of the Hamiltonian $\mathcal{H}_{\tau_z}(\bm{k})=v_F(\tau_z k_x\sigma_x+k_y\sigma_y)+m\sigma_z$ with 
eigenvalues $\varepsilon^\pm_{\bm{k}}=\pm\varepsilon_{\bm{k}}=\pm\sqrt{v_F^2(k_x^2+k_y^2)+m^2}$ are given by
\begin{align}
|u_{\bm{k}}^\pm(\tau_z)\rangle&=\frac{1}{\sqrt{2}}
\begin{bmatrix}
\sqrt{1\pm\frac{m}{\varepsilon_{\bm{k}}}}\\
\pm \tau_z e^{i\tau_z\theta}\sqrt{1\mp\frac{m}{\varepsilon_{\bm{k}}}}
\end{bmatrix},
\end{align}
where $e^{\pm i\theta}=(k_x\pm ik_y)/k$ with $k=\sqrt{k_x^2+k_y^2}$.
From these eigenstates, we immediately get
\begin{align}
U^{++}_{\bm{k}\bm{k}'}=U\langle u^+_{\bm{k}}(\tau_z) | u^+_{\bm{k'}}(\tau_z)\rangle=\frac{U_0}{2}\left[\sqrt{\left(1+\frac{m}{\varepsilon}\right)\left(1+\frac{m'}{\varepsilon'}\right)}+e^{i\tau_z\gamma}\sqrt{\left(1-\frac{m}{\varepsilon}\right)\left(1-\frac{m'}{\varepsilon'}\right)}\right],\nonumber \\
U^{+-}_{\bm{k}\bm{k}'}=U\langle u^+_{\bm{k}}(\tau_z) | u^-_{\bm{k'}}(\tau_z)\rangle=\frac{U_0}{2}\left[\sqrt{\left(1+\frac{m}{\varepsilon}\right)\left(1-\frac{m'}{\varepsilon'}\right)}-e^{i\tau_z\gamma}\sqrt{\left(1-\frac{m}{\varepsilon}\right)\left(1+\frac{m'}{\varepsilon'}\right)}\right],\nonumber \\
U^{-+}_{\bm{k}\bm{k}'}=U\langle u^-_{\bm{k}}(\tau_z) | u^+_{\bm{k'}}(\tau_z)\rangle=\frac{U_0}{2}\left[\sqrt{\left(1-\frac{m}{\varepsilon}\right)\left(1+\frac{m'}{\varepsilon'}\right)}-e^{i\tau_z\gamma}\sqrt{\left(1+\frac{m}{\varepsilon}\right)\left(1-\frac{m'}{\varepsilon'}\right)}\right],\nonumber \\
U^{--}_{\bm{k}\bm{k}'}=U\langle u^-_{\bm{k}}(\tau_z) | u^-_{\bm{k'}}(\tau_z)\rangle=\frac{U_0}{2}\left[\sqrt{\left(1-\frac{m}{\varepsilon}\right)\left(1-\frac{m'}{\varepsilon'}\right)}+e^{i\tau_z\gamma}\sqrt{\left(1+\frac{m}{\varepsilon}\right)\left(1+\frac{m'}{\varepsilon'}\right)}\right],
\label{U^mm}
\end{align}
where $\gamma=\theta'-\theta$.
Then we obtain
\begin{align}
\langle U^{++}_{\bm{k}\bm{k}'}U^{+-}_{\bm{k}'\bm{k}}\rangle&=\frac{n_{\mathrm{imp}}U_0^2}{4}\left[\sqrt{\left(1+\frac{m}{\varepsilon}\right)\left(1+\frac{m'}{\varepsilon'}\right)}+e^{i\tau_z\gamma}\sqrt{\left(1-\frac{m}{\varepsilon}\right)\left(1-\frac{m'}{\varepsilon'}\right)}\right]\left[\sqrt{\left(1+\frac{m'}{\varepsilon'}\right)\left(1-\frac{m}{\varepsilon}\right)}-e^{-i\tau_z\gamma}\sqrt{\left(1-\frac{m'}{\varepsilon'}\right)\left(1+\frac{m}{\varepsilon}\right)}\right]\nonumber \\
&=\frac{n_{\mathrm{imp}}U_0^2}{2}\left[\frac{v_F k}{\varepsilon}\frac{m'}{\varepsilon'}+\left(i\tau_z\sin\gamma-\frac{m}{\varepsilon}\cos\gamma\right)\frac{v_F k'}{\varepsilon'}\right],\nonumber \\
\langle U^{+-}_{\bm{k}\bm{k}'}U^{--}_{\bm{k}'\bm{k}}\rangle&=\frac{n_{\mathrm{imp}}U_0^2}{4}\left[\sqrt{\left(1+\frac{m}{\varepsilon}\right)\left(1-\frac{m'}{\varepsilon'}\right)}-e^{i\tau_z\gamma}\sqrt{\left(1-\frac{m}{\varepsilon}\right)\left(1+\frac{m'}{\varepsilon'}\right)}\right]\left[\sqrt{\left(1-\frac{m'}{\varepsilon'}\right)\left(1-\frac{m}{\varepsilon}\right)}+e^{-i\tau_z\gamma}\sqrt{\left(1+\frac{m'}{\varepsilon'}\right)\left(1+\frac{m}{\varepsilon}\right)}\right]\nonumber \\
&=\frac{n_{\mathrm{imp}}U_0^2}{2}\left[-\frac{v_F k}{\varepsilon}\frac{m'}{\varepsilon'}+\left(-i\tau_z\sin\gamma+\frac{m}{\varepsilon}\cos\gamma\right)\frac{v_F k'}{\varepsilon'}\right]\nonumber \\
&=-\langle U^{++}_{\bm{k}\bm{k}'}U^{+-}_{\bm{k}'\bm{k}}\rangle,\nonumber \\
\langle U^{--}_{\bm{k}\bm{k}'}U^{-+}_{\bm{k}'\bm{k}}\rangle&=\langle U^{++}_{\bm{k}\bm{k}'}U^{+-}_{\bm{k}'\bm{k}}\rangle\ (m\rightarrow -m,\ m'\rightarrow -m')=-\left[\langle U^{++}_{\bm{k}\bm{k}'}U^{+-}_{\bm{k}'\bm{k}}\rangle\right]^*,\nonumber \\
\langle U^{-+}_{\bm{k}\bm{k}'}U^{++}_{\bm{k}'\bm{k}}\rangle&=\langle U^{+-}_{\bm{k}\bm{k}'}U^{--}_{\bm{k}'\bm{k}}\rangle\ (m\rightarrow -m,\ m'\rightarrow -m')=\left[\langle U^{++}_{\bm{k}\bm{k}'}U^{+-}_{\bm{k}'\bm{k}}\rangle\right]^*,
\end{align}

From the definition of $J(\langle n\rangle)_{\bm{k}}^{+-}$ with $\langle n\rangle=\mathrm{diag}[n^+_{\bm{k}}, n^-_{\bm{k}}]$, we have
\begin{align}
[J(\langle n\rangle)]_{\bm{k}}^{+-}&=\pi\sum_{\bm{k}'}\langle U^{++}_{\bm{k}\bm{k}'}U^{+-}_{\bm{k}'\bm{k}}\rangle\left[(n_{\bm{k}}^+-n_{\bm{k}'}^+)\delta(\varepsilon^+_{\bm{k}}-\varepsilon^+_{\bm{k}'})+(n_{\bm{k}}^--n_{\bm{k}'}^+)\delta(\varepsilon^-_{\bm{k}}-\varepsilon^+_{\bm{k}'})\right]\nonumber \\
&\quad+\pi\sum_{\bm{k}'}\langle U^{+-}_{\bm{k}\bm{k}'}U^{--}_{\bm{k}'\bm{k}}\rangle\left[(n_{\bm{k}}^+-n_{\bm{k}'}^-)\delta(\varepsilon^+_{\bm{k}}-\varepsilon^-_{\bm{k}'})+(n_{\bm{k}}^--n_{\bm{k}'}^-)\delta(\varepsilon^-_{\bm{k}}-\varepsilon^-_{\bm{k}'})\right]\nonumber \\
&=\pi\sum_{\bm{k}'}\langle U^{++}_{\bm{k}\bm{k}'}U^{+-}_{\bm{k}'\bm{k}}\rangle\left[(n_{\bm{k}}^+-n_{\bm{k}'}^+)\delta(\varepsilon^+_{\bm{k}}-\varepsilon^+_{\bm{k}'})-(n_{\bm{k}}^--n_{\bm{k}'}^-)\delta(\varepsilon^-_{\bm{k}}-\varepsilon^-_{\bm{k}'})\right],
\label{J^+-}
\end{align}
where we have used $\delta(\varepsilon^-_{\bm{k}}-\varepsilon^+_{\bm{k}'})=\delta(\varepsilon^+_{\bm{k}}-\varepsilon^-_{\bm{k}'})=0$.
Similarly, we get
\begin{align}
[J(\langle n\rangle)]_{\bm{k}}^{-+}&=\pi\sum_{\bm{k}'}\langle U^{-+}_{\bm{k}\bm{k}'}U^{++}_{\bm{k}'\bm{k}}\rangle\left[(n_{\bm{k}}^--n_{\bm{k}'}^+)\delta(\varepsilon^-_{\bm{k}}-\varepsilon^+_{\bm{k}'})+(n_{\bm{k}}^+-n_{\bm{k}'}^+)\delta(\varepsilon^+_{\bm{k}}-\varepsilon^+_{\bm{k}'})\right]\nonumber \\
&\quad+\pi\sum_{\bm{k}'}\langle U^{--}_{\bm{k}\bm{k}'}U^{-+}_{\bm{k}'\bm{k}}\rangle\left[(n_{\bm{k}}^--n_{\bm{k}'}^-)\delta(\varepsilon^-_{\bm{k}}-\varepsilon^-_{\bm{k}'})+(n_{\bm{k}}^+-n_{\bm{k}'}^-)\delta(\varepsilon^+_{\bm{k}}-\varepsilon^-_{\bm{k}'})\right]\nonumber \\
&=\pi\sum_{\bm{k}'}\langle U^{++}_{\bm{k}\bm{k}'}U^{+-}_{\bm{k}'\bm{k}}\rangle^*\left[(n_{\bm{k}}^+-n_{\bm{k}'}^+)\delta(\varepsilon^+_{\bm{k}}-\varepsilon^+_{\bm{k}'})-(n_{\bm{k}}^--n_{\bm{k}'}^-)\delta(\varepsilon^-_{\bm{k}}-\varepsilon^-_{\bm{k}'})\right]=\left[J(\langle n\rangle)^{+-}\right]^*.
\label{J^-+}
\end{align}

\subsection{Correction to the linear magnetoconductivity due to disorder scattering}
Let us calculate the contribution from $J(\langle n\rangle)$ to the linear magnetoconductivity, which corresponds to the vertex correction in the ladder-diagram approximation.
The correction to $\langle S_{B}\rangle$ [Eq.~(\ref{S_EB})] is given by
\begin{align}
\langle S'_{B}\rangle_{\bm{k}}^{nn'}=i\frac{[J(\langle n_{B}\rangle)]_{\bm{k}}^{nn'}}{\varepsilon_{\bm{k}}^n-\varepsilon_{\bm{k}}^{n'}}
=
\begin{bmatrix}
0 && i\frac{J^{+-}}{\varepsilon^+_{\bm{k}}-\varepsilon^-_{\bm{k}}}\\
-i\frac{(J^{+-})^*}{\varepsilon^+_{\bm{k}}-\varepsilon^-_{\bm{k}}} && 0
\end{bmatrix}
=-\tilde{\sigma}_y\frac{\mathrm{Re}\left\{[J(\langle n_B\rangle)]_{\bm{k}}^{+-}\right\}}{2\varepsilon_{\bm{k}}}-\tilde{\sigma}_x\frac{\mathrm{Im}\left\{[J(\langle n_B\rangle)_{\bm{k}}^{+-}\right\}}{2\varepsilon_{\bm{k}}},
\end{align}
where
\begin{align}
\mathrm{Re}\left\{[J(\langle n_B\rangle)]_{\bm{k}}^{+-}\right\}&=\pi\frac{n_{\mathrm{imp}}U_0^2}{2}\sum_{\bm{k}'}\left[\frac{v_F k}{\varepsilon}\frac{m'}{\varepsilon'}-\frac{m}{\varepsilon}\frac{v_F k'}{\varepsilon'}(\cos\theta'\cos\theta+\sin\theta\sin\theta')\right](n_{B\bm{k}}^+-n_{B\bm{k}'}^+)\delta(\varepsilon^+_{\bm{k}}-\varepsilon^+_{\bm{k}'})\nonumber\\
&=\pi\frac{n_{\mathrm{imp}}U_0^2}{2}\frac{v_F k}{\varepsilon}n_{B\bm{k}}^+\sum_{\bm{k}'}\frac{m'}{\varepsilon'}\delta(\varepsilon^+_{\bm{k}}-\varepsilon^+_{\bm{k}'})+\pi\frac{n_{\mathrm{imp}}U_0^2}{2}\frac{m}{\varepsilon}\sin\theta\sum_{\bm{k}'}\frac{v_F k'}{\varepsilon'}\sin\theta' n_{B\bm{k}'}^+\delta(\varepsilon^+_{\bm{k}}-\varepsilon^+_{\bm{k}'}),
\end{align}
and
\begin{align}
\mathrm{Im}\left\{[J(\langle n_B\rangle)]_{\bm{k}}^{+-}\right\}&=\tau_z\pi\frac{n_{\mathrm{imp}}U_0^2}{2}\sum_{\bm{k}'}(\sin\theta'\cos\theta-\cos\theta'\sin\theta)\frac{v_F k'}{\varepsilon'}(n_{B\bm{k}}^+-n_{B\bm{k}'}^+)\delta(\varepsilon^+_{\bm{k}}-\varepsilon^+_{\bm{k}'})\nonumber\\
&=-\tau_z\pi\frac{n_{\mathrm{imp}}U_0^2}{2}\cos\theta\sum_{\bm{k}'}\sin\theta'\frac{v_F k'}{\varepsilon'}n_{B\bm{k}'}^+\delta(\varepsilon^+_{\bm{k}}-\varepsilon^+_{\bm{k}'}),
\end{align}
with
\begin{align}
n_{B\bm{k}}^+=\tau_{\mathrm{tr}}eB_z\frac{\partial \varepsilon_{\bm{k}}}{\partial k_y}\frac{\partial n^+_{E\bm{k}}}{\partial k_x}-\frac{\partial \varepsilon_{\bm{k}}}{\partial k_x}\frac{\partial n^+_{E\bm{k}}}{\partial k_y}=\tau_{\mathrm{tr}}^2e^2E_xB_z\left(\frac{\partial \varepsilon_{\bm{k}}}{\partial k_y}\frac{\partial }{\partial k_x}-\frac{\partial \varepsilon_{\bm{k}}}{\partial k_x}\frac{\partial }{\partial k_y}\right)\frac{\partial f_0(\varepsilon_{\bm{k}}^+)}{\partial k_x},
\end{align}
and $n^+_{E\bm{k}}=\tau_{\mathrm{tr}}eE_x\frac{\partial f_0(\varepsilon_{\bm{k}}^+)}{\partial k_x}$ [see Eq.~(\ref{n_EB})].
Here, we have used the fact that $n_{B\bm{k}}^+$ is an odd function of $k_y$, and $\sin\gamma=\sin\theta'\cos\theta-\cos\theta'\sin\theta$ and $\cos\gamma=\cos\theta'\cos\theta+\sin\theta\sin\theta'$.
Then we have
\begin{align}
v_x \langle S'_{B}\rangle&=v_F\left(\tilde{\sigma}_z\frac{v_F k}{\varepsilon_{\bm{k}}}\cos\theta+\tilde{\sigma}_y\tau_z\sin\theta-\tilde{\sigma}_x\frac{m}{\varepsilon_{\bm{k}}}\cos\theta\right)
\left[-\tilde{\sigma}_y\frac{\mathrm{Re}\left\{[J(\langle n_B\rangle)]_{\bm{k}}^{+-}\right\}}{2\varepsilon_{\bm{k}}}-\tilde{\sigma}_x\frac{\mathrm{Im}\left\{[J(\langle n_B\rangle)_{\bm{k}}^{+-}\right\}}{2\varepsilon_{\bm{k}}}\right]\nonumber\\
&=\tau_z v_F\pi\frac{n_{\mathrm{imp}}U_0^2}{2}\left[-\frac{m}{2\varepsilon_{\bm{k}}^2}\sum_{\bm{k}'}\sin\theta'\frac{v_F k'}{\varepsilon'}n_{B\bm{k}'}^+\delta(\varepsilon^+_{\bm{k}}-\varepsilon^+_{\bm{k}'})
-\frac{v_F k}{2\varepsilon^2}\sin\theta n_{B\bm{k}}^+\sum_{\bm{k}'}\frac{m'}{\varepsilon'}\delta(\varepsilon^+_{\bm{k}}-\varepsilon^+_{\bm{k}'})\right]\bm{1}+\mathrm{(traceless\ terms)}.
\end{align}
Finally, we obtain the valley-dependent vertex correction due to disorder scattering to the linear magnetoconductivity
\begin{align}
\sigma_{xx}^{(1)\mathrm{ver}}(B_z)&=\mathrm{Tr}\left[(-e) v_x \langle S'_{B}\rangle\right]/E_x\nonumber\\
&=(2e)\tau_z v_F\pi\frac{n_{\mathrm{imp}}U_0^2}{2}\int_{-\infty}^\infty\frac{d^2k}{(2\pi)^2}\int_{-\infty}^\infty\frac{d^2k'}{(2\pi)^2}\delta(\varepsilon^+_{\bm{k}}-\varepsilon^+_{\bm{k}'})\left[\frac{m}{2\varepsilon_{\bm{k}}^2}\sin\theta'\frac{v_F k'}{\varepsilon'}n_{B\bm{k}'}^+
+\frac{v_F k}{2\varepsilon^2}\sin\theta n_{B\bm{k}}^+\frac{m'}{\varepsilon'}\right]\nonumber\\
&\equiv\tau_z \frac{e^3}{\hbar}B_z v_F^2\mathcal{C}_1^{\mathrm{dis}}(\mu,m)\tau_{\mathrm{tr}},
\end{align}
where $\mathcal{C}_1^{\mathrm{dis}}(\mu,m)<0$ and $\tau_{\mathrm{tr}}=(4\hbar v_F^2/n_{\mathrm{imp}}U_0^2\mu)(1+3m^2/\mu^2)^{-1}$ is the intravalley scattering time (see the next section for a detailed calculation).
Note that $n_{B\bm{k}}^+\propto \tau_{\mathrm{tr}}^2$, while the coefficient in front of the integral is proportional to $\tau_{\mathrm{tr}}^{-1}$.

\subsection{Calculation of the intravalley scattering time}
Let us calculate the intravalley scattering time (or equivalently transport lifetime) in the two-band massive Dirac model we have studied.
The intravalley scattering time of band $m$ can be calculated from the following well-known equation:
\begin{align} 
\frac{1}{\tau_{\mathrm{tr}}^m}=\frac{2\pi}{\hbar}\int\frac{d^2k'}{(2\pi)^2}\, \langle |U^{mm}_{\bm{k}\bm{k}'}|^2 
\rangle(1-\cos\theta_{\bm{k}\bm{k}'}) \delta(\mu - \varepsilon^m_{\bm{k}'}),
\end{align}
where $\cos\theta_{\bm{k}\bm{k}'}=\bm{k}\cdot\bm{k}'/|\bm{k}||\bm{k}'|$.
Because $\langle |U^{++}_{\bm{k}\bm{k}'}|^2 \rangle=\langle |U^{--}_{\bm{k}\bm{k}'}|^2 \rangle$, we may drop the band index and consider the case of $\mu>0$.
From Eq.~(\ref{U^mm}) we get
\begin{align} 
\langle |U^{++}_{\bm{k}\bm{k}'}|^2 \rangle&=\frac{n_{\mathrm{imp}}U_0^2}{4}\left|\left[\sqrt{\left(1+\frac{m}{\varepsilon}\right)\left(1+\frac{m'}{\varepsilon'}\right)}+(\cos\gamma+i\tau_z\sin\gamma)\sqrt{\left(1-\frac{m}{\varepsilon}\right)\left(1-\frac{m'}{\varepsilon'}\right)}\right]\right|^2\nonumber \\
&=\frac{n_{\mathrm{imp}}U_0^2}{2}\left[1+\frac{m^2}{\varepsilon_{\bm{k}}\varepsilon_{\bm{k}'}}+\cos\gamma\frac{v_F^2 k k'}{\varepsilon_{\bm{k}}\varepsilon_{\bm{k}'}}\right].
\end{align}
We define the Fermi wave number $k=k'=k_F\equiv \sqrt{\mu^2-m^2}/v_F$ ($\varepsilon_{\bm{k}}=\varepsilon_{\bm{k}'}=\mu$) and $\theta_{\bm{k}\bm{k}'}\equiv \phi$.
Also, without loss of generality we can set $\bm{k}=(k_F,0)$ (i.e., $\theta=0$), which leads to $\cos\gamma=\cos\phi$.
Finally, we obtain
\begin{align} 
\frac{1}{\tau_{\mathrm{tr}}}&=\frac{2\pi}{\hbar}\frac{n_{\mathrm{imp}}U_0^2}{2}\frac{1}{4\pi^2}\int_0^{\infty} k'dk' \int_0^{2\pi}d\phi\left[1+\frac{m^2}{\varepsilon_{\bm{k}}\varepsilon_{\bm{k}'}}+\cos\gamma\frac{v_F^2 k k'}{\varepsilon_{\bm{k}}\varepsilon_{\bm{k}'}}\right]\delta(\mu - \varepsilon^m_{\bm{k}'}),\nonumber \\
&=\frac{n_{\mathrm{imp}}U_0^2\mu}{4\hbar v_F^2}\left(1+3\frac{m^2}{\mu^2}\right).
\end{align}

\section{Approximate forms of the longitudinal magnetoresistance $\Delta\rho_{xx}$}
Let us derive approximate forms of the longitudinal magnetoresistance $\Delta\rho_{xx}(B_z)\equiv[\rho_{xx}(B_z)-\rho_{xx}(0)]/\rho_{xx}(0)$, where $\rho_{xx}(B_z)=\sigma_{xx}/(\sigma_{xx}^2+\sigma_{xy}^2)$ with $\sigma_{\mu\nu}\equiv\sigma_{\mu\nu}^{(0)}+\sigma_{\mu\nu}^{B}$, in the low-field limit with the condition $\sigma^{(0)}_{xx}\gg\sigma^{(0)}_{xy}$ or $\sigma^{(0)}_{xx}\ll\sigma^{(0)}_{xy}$ for the massive Dirac model (\ref{MassiveDirac}).
In this model, the anomalous Hall conductivity $\sigma^{(0)}_{xy}$ takes the maximum value $\tau_z e^2/2h$ when the Fermi level lies in the gap.

First, let us consider the case of $\sigma^{(0)}_{xx}\gg\sigma^{(0)}_{xy}$, i.e., the case of high-carrier-desity semiconductors ($\mu>m$).
In this case, we have
\begin{align}
\rho_{xx}(B_z)&=\frac{\sigma^{(0)}_{xx}+\sigma_{xx}^{B}}{(\sigma^{(0)}_{xx}+\sigma_{xx}^{B})^2+(\sigma^{(0)}_{xy}+\sigma_{xy}^{B})^2}
\approx \frac{\sigma^{(0)}_{xx}+\sigma_{xx}^{B}}{(\sigma^{(0)}_{xx})^2}\left[1-2\frac{\sigma^{(0)}_{xx}\sigma^{B}_{xx}+\sigma^{(0)}_{xy}\sigma^{B}_{xy}}{(\sigma^{(0)}_{xx})^2}\right]\approx \frac{\sigma^{(0)}_{xx}+\sigma_{xx}^{B}}{(\sigma^{(0)}_{xx})^2}-2\frac{\sigma^{(0)}_{xx}\sigma^{B}_{xx}+\sigma^{(0)}_{xy}\sigma^{B}_{xy}}{(\sigma^{(0)}_{xx})^3} \nonumber\\
&\approx \frac{\sigma^{(0)}_{xx}-\sigma^{B}_{xx}}{(\sigma^{(0)}_{xx})^2},
\end{align}
where we have used that $(\sigma^{B}_{\mu\nu}/\sigma^{(0)}_{xx})^2\ll 1$ and $(\sigma^{(0)}_{xy}/\sigma^{(0)}_{xx})(\sigma^{B}_{xy}/\sigma^{(0)}_{xx})\ll 1$.
We also have $\rho_{xx}(0)=\frac{\sigma^{(0)}_{xx}}{(\sigma^{(0)}_{xx})^2+(\sigma^{(0)}_{xy})^2}\approx 1/\sigma^{(0)}_{xx}$.
Then the longitudinal magnetoresistance $\Delta\rho_{xx}(B_z)$ in the low-field limit is obtained as
\begin{align} 
\Delta\rho_{xx}(B_z)\approx -\frac{\sigma^{B}_{xx}}{\sigma^{(0)}_{xx}}.
\end{align}

Next, let us consider the case of $\sigma^{(0)}_{xx}\ll\sigma^{(0)}_{xy}$, i.e., the case of low-carrier-density semiconductors (with small $m$) or the case of $\mu\approx m$.
In this case, we have
\begin{align}
\rho_{xx}(B_z)&=\frac{\sigma^{(0)}_{xx}+\sigma_{xx}^{B}}{(\sigma^{(0)}_{xx}+\sigma_{xx}^{B})^2+(\sigma^{(0)}_{xy}+\sigma_{xy}^{B})^2}
\approx \frac{\sigma^{(0)}_{xx}+\sigma_{xx}^{B}}{(\sigma^{(0)}_{xy})^2}\left[1-2\frac{\sigma^{(0)}_{xx}\sigma^{B}_{xx}+\sigma^{(0)}_{xy}\sigma^{B}_{xy}}{(\sigma^{(0)}_{xy})^2}\right]\approx \frac{\sigma^{(0)}_{xx}+\sigma_{xx}^{B}}{(\sigma^{(0)}_{xy})^2}-2\sigma^{(0)}_{xx}\frac{\sigma^{(0)}_{xx}\sigma^{B}_{xx}+\sigma^{(0)}_{xy}\sigma^{B}_{xy}}{(\sigma^{(0)}_{xy})^4}\nonumber\\
&\approx \frac{\sigma^{(0)}_{xx}+\sigma_{xx}^{B}}{(\sigma^{(0)}_{xy})^2},
\end{align}
where we have used that $(\sigma^{B}_{\mu\nu}/\sigma^{(0)}_{xy})^2\ll 1$ and $(\sigma^{(0)}_{xx}/\sigma^{(0)}_{xy})(\sigma^{B}_{xy}/\sigma^{(0)}_{xy})\ll 1$.
We also have $\rho_{xx}(0)=\frac{\sigma^{(0)}_{xx}}{(\sigma^{(0)}_{xx})^2+(\sigma^{(0)}_{xy})^2}\approx \sigma^{(0)}_{xx}/(\sigma^{(0)}_{xy})^2$.
Then the longitudinal magnetoresistance $\Delta\rho_{xx}(B_z)$ in the low-field limit is obtained as
\begin{align} 
\Delta\rho_{xx}(B_z)\approx +\frac{\sigma^{B}_{xx}}{\sigma^{(0)}_{xx}}.
\end{align}

\end{widetext}


\begin{thebibliography}{999}
\bibitem{Comment6}
The quantum kinetic equation~(\ref{full-kinetic-equation}) reduces to the usual semiclassical Boltzmann equation in spin-independent single-band systems, since the covariant derivatives reduce to simple derivatives in such systems. In this work we treat disorder in the Born approximation, although we can in principle go beyond this in our formalism. In the approximation we employ, the results obtained by our quantum kinetic formalism are equivalent to those obtained by the Kubo formalism with vertex corrections included in the ladder-diagram approximation and both band diagonal and band off-diagonal terms taken into account.  If only band diagonal terms were retained, our approach would be equivalent to the standard semiclassical Boltzmann theory.  One of the advantages of our formalism is that Berry phase and disorder contributions can be incorporated in a unified fashion at each order of electric and magnetic field strengths.
\bibitem{Comment2}
Under time reversal valleys change place and the field direction is reversed.
The response must be odd in time and odd in field or even in time and even in field.
\bibitem{Comment4}
Note that Fig.~\ref{Fig1}(b) is an enlarged view of the absolute values of the magnetoresistance $\Delta\rho_{xx}$ at the magnetic field scale $B_z\sim 0.01\, \mathrm{T}$.
\end{thebibliography}
\end{document}